\documentclass[12pt, letterpaper, preprint, tighten]{aastex62}
\usepackage{scalerel}
\usepackage{amsmath}
\usepackage{color}
\usepackage{comment}
\usepackage{makecell}

\let\oldbibliography\thebibliography 
\renewcommand{\thebibliography}[1]{%
  \oldbibliography{#1}%
  \setlength{\itemsep}{0pt}%
  \setlength{\parsep}{0pt}%
  \setlength{\parskip}{0pt}%
  \setlength{\bibsep}{0ex}
  \raggedright
}
\definecolor{orange}{rgb}{1,0.5,0}
\definecolor{dred}{rgb}{0.59,0.,0.09}

\newcommand{\beq}{\begin{equation}}
\newcommand{\eeq}{\end{equation}}

\newcommand{\cao}{Cao et al. (in preparation)}

\newcommand{\tduty}{t_{\rm duty}}
\newcommand{\logsfr}{\log\mathrm{SFR}}
\newcommand{\logsfrsfs}{\log\mathrm{SFR}_\mathrm{SFS}}
\newcommand{\bitem}{\begin{itemize}}
\newcommand{\eitem}{\end{itemize}}

\newcommand{\siglogm}{\sigma_{\scaleto{M_*|M_h}{8pt}}}
\newcommand{\sigtwe}{\sigma_{\scaleto{M_*|M_h=10^{12}M_\odot}{8pt}}}

\begin{document}\sloppy\sloppypar\frenchspacing

\title{Constraining Star Formation Histories of Blue Galaxies using the Scatter between Stellar Mass and Halo Mass} 
\author{ChangHoon Hahn}
\altaffiliation{hahn.changhoon@gmail.com}
\affil{Lawrence Berkeley National Laboratory, 1 Cyclotron Rd, Berkeley CA 94720, USA}
\affil{Berkeley Center for Cosmological Physics, University of California, Berkeley, CA 94720, USA}
\affil{Center for Cosmology and Particle Physics, Department of Physics, New York University, 4 Washington Place, New York, NY 10003}
\author{Jeremy L.~Tinker}
\affil{Center for Cosmology and Particle Physics, Department of Physics, New York University, 4 Washington Place, New York, NY 10003}
\author{Andrew Wetzel}
\affil{Department of Physics, University of California, Davis, CA USA}

\begin{abstract}
    We present constraints on the timescale of star formation variability and 
    the correlation between star formation and host halo accretion histories 
    of star-forming (SF) central galaxies from the measured scatter of the stellar-to-halo 
    mass relation (SHMR). SF galaxies are found to have a tight relationship
    between their star formation rates and stellar masses on the so-called ``star-forming
    sequence'' (SFS), which characterizes both their star formation histories and
    stellar mass growths. Meanwhile, observed constraints on the SHMR connect 
    stellar mass growth to host halo accretion history. Combining these observed 
    trends with a cosmological $N$-body simulation, we present flexible models that 
    track the star formation, stellar mass, and host halo accretion histories of 
    SF central galaxies at $z < 1$ while reproducing the observed stellar 
    mass function and SFS of central galaxies in SDSS Data Release 7. Using these 
    models, we find that the scatter in SHMR at $M_h{=}10^{12}M_\odot$, $\sigtwe$, 
    is sensitive to the timescale of star 
    formation variability, $t_{\rm duty}$, and the correlation coefficient, $r$, 
    between star formation and host halo accretion histories: shorter $\tduty$ and 
    higher $r$ both result in tighter $\sigtwe$. To reproduce a constant 
    $\siglogm \sim 0.2$ dex over $z=1$ to 0, our models require $\tduty \leq 1.5$ Gyr 
    for $r = 0.99$ or $r > 0.6$ for $\tduty = 0.1$ Gyr. For $r \sim 0.6$, as found 
    in the literature, $\tduty < 0.2$ Gyr is necessary. Meanwhile, to reproduce 
    the tightening of $\sigtwe = 0.35$ to 0.2 dex from $z=1$ to 0 in hydrodynamical 
    simulations, our models require $\tduty = 0.1$ Gyr for $r > 0.5$. 
    Although, the lack of consensus on $\siglogm$ at $M_h=10^{12}M_\odot$ and at 
    $z=1$ from observations and galaxy formation models remains the main bottleneck 
    in precisely constraining $r$ and $t_\mathrm{duty}$, we demonstrate that SHMR 
    can be used to the constrain star formation and host halo accretion histories of 
    SF central galaxies.
\end{abstract}
\keywords{methods: numerical -- galaxies: evolution -- galaxies: haloes --
galaxies: star formation -- galaxies: groups: general -- cosmology: observations.}

\section{Introduction}
\begin{figure}
\begin{center}
\includegraphics[width=0.75\textwidth]{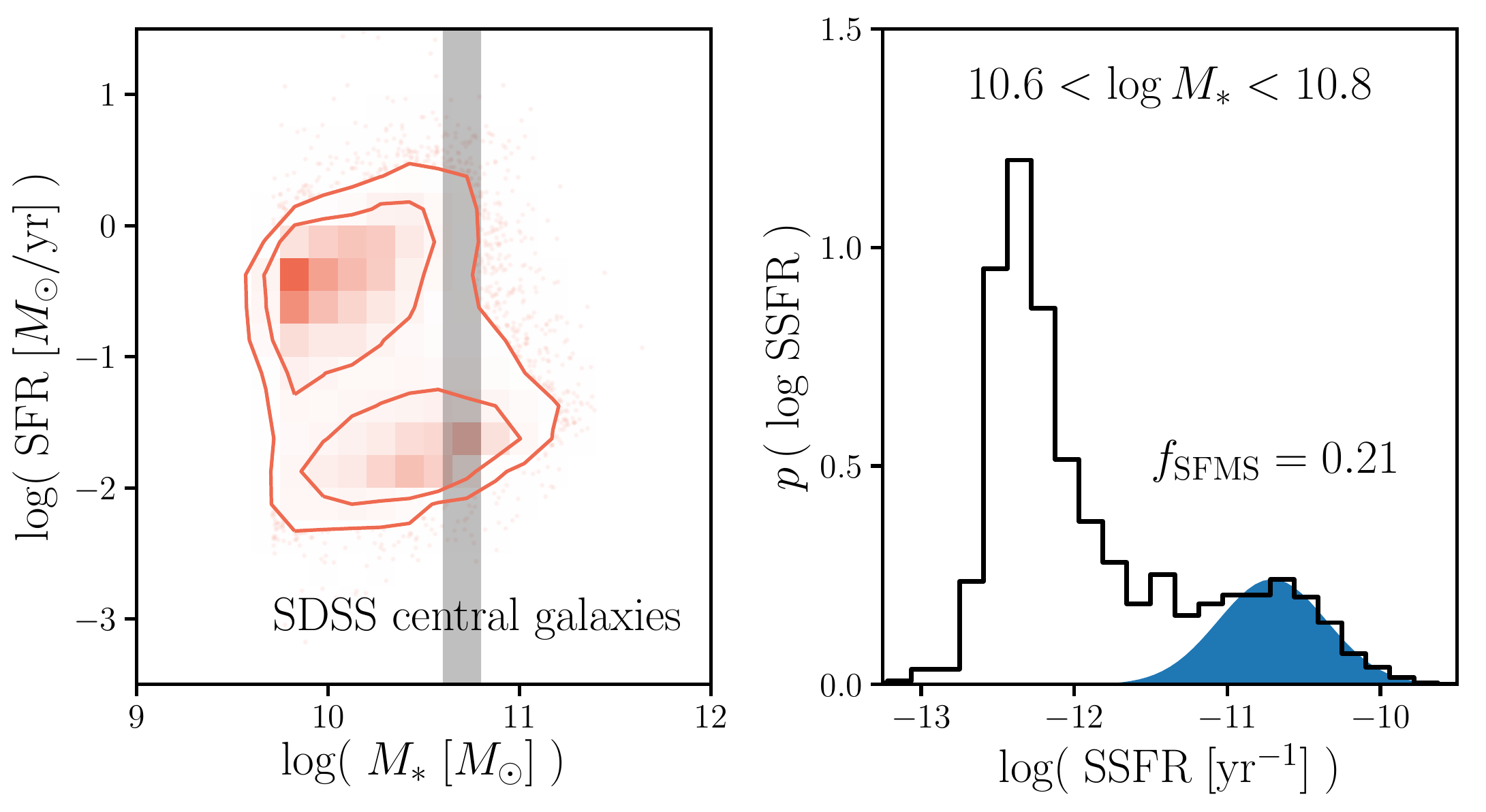}
    \caption{The SFR--$M_*$ relation of the central galaxies in SDSS DR7
    mark the bimodal distribution of the SF and quiescent populations (left panel). 
    \emph{SF centrals, based on the correlation between their
    SFR and $M_*$, lie on the so-called ``star-forming sequence''}.
    On the right, we present the SSFR distribution, $p(\log\mathrm{SSFR})$,
    of SDSS centrals with $10.6 < \log M_* < 10.8$. Based on the SFS component
    from the \cite{hahn2018a} GMM fit to the SFR--$M_*$ relation (shaded in blue),
    galaxies in the SFS account for $f_\mathrm{SFS} = 0.21$ of the centrals
    in the stellar mass bin.} \label{fig:groupcat}
\end{center}
\end{figure}

Observations from large surveys such as the Sloan Digital Sky Survey~\citep[SDSS;][]{york2000}
have been critical for establishing the global trends of galaxies in
the local universe. Broadly speaking, galaxies fall into two categories:
quiescent and star-forming (hereafter SF) galaxies. Quiescent galaxies
have little to no star formation, are red in color due to old stellar populations,
and have elliptical morphologies. Meanwhile, SF galaxies have significant star 
formation, thus are blue in color, and have disk-like morphologies 
(\citealt{kauffmann2003, blanton2003, baldry2006, taylor2009, moustakas2013}; 
see~\citealt{blanton2009} and references therein). SF galaxies, furthermore, are 
found on the so-called ``star-forming sequence'' (hereafter SFS), a tight 
relationship between their star formation rates (SFR) and stellar 
masses~\citep[][see also Figure~\ref{fig:groupcat}]{noeske2007, daddi2007, salim2007, speagle2014, lee2015}.
This sequence, which is observed out to $z > 2$~\citep{wang2013, leja2015}
plays a crucial role in determining galaxy evolution over the past 
${\sim}10\,\mathrm{Gyr}$~\citep[see][for an alternative point of view]{kelson2014,abramson2016}.
The significant fraction of SF galaxies that quench their
star formation and migrate off of the SFS reflects the growth in the
fraction of quiescent galaxies~\citep{blanton2006, borch2006, bundy2006, moustakas2013}.
The decline of star formation in the entire SFS~\citep{lee2015, schreiber2015}
over time reflects the decline in overall cosmic star formation~\citep{hopkins2006, behroozi2013, madau2014}.
With its evolution, the SFS also connects the star formation histories of SF
galaxies to their stellar mass growths.

Recent observations have also allowed us to investigate how galaxies fit
into the context of hierarchical structure formation predicted by 
$\Lambda$CDM cosmology. In addition to traditional theoretical
models of hydrodynamic simulations and semi-analytic models~(see \citealt{silk2012, somerville2015} for reviews),
empirical models have been remarkably effective for understanding the 
galaxy-halo connection. These models relate galaxy properties to their 
host dark matter halo properties using methods such as halo occupation 
distribution modeling~\citep[HOD; \emph{e.g.}][]{zheng2007,zehavi2011,leauthaud2012,parejko2013,zu2015},
conditional luminosity function modeling~\citep[\emph{e.g.}][]{yang2009}, 
and abundance matching~\citep[\emph{e.g.}][]{kravtsov2004, vale2006, conroy2009, moster2013, reddick2013}.
Using these models, more massive halos are found to host more massive 
galaxies on the stellar-to-halo mass relation~\citep[hereafter SHMR;][]{mandelbaum2006a, conroy2007, more2011, leauthaud2012, tinker2013, velander2014, han2015, zu2015, gu2016, lange2018a} 
with a tight scatter in $\log M_*$ at fixed $M_h$ --- $\siglogm$ --- of $0.2$ dex. 
These constraints are mainly driven by massive halos with $M_h > 10^{12} M_\odot$. 
A similarly tight scatter is found at higher $z\sim1$~\citep{leauthaud2012, tinker2013, patel2015}.
The tight scatter in SHMR over $z < 1$ suggests that stellar mass growth of 
galaxies is linked to the growth of their host dark matter halos.

Despite these developments, we face a number of challenges when it comes
to understanding the detailed star formation histories (SFH) and its
connection to host halo assembly history of galaxies.
For instance, SFHs at lookback times longer than $200\,\mathrm{Myr}$
do not contribute to SFR indicators such as $H\alpha$ or $FUV$ fluxes~\citep{sparre2017}.
Measuring SFHs from fitting photometry or spectroscopy typically
assume a specific functional form of the SFH, such as exponentially
declining or lognormal, that do not include variations on short
timescales~\citep[\emph{e.g.}][]{wilkinson2017, carnall2018}.
Even methods that recover non-parametric SFHs from high signal-to-noise
observations can only retrieve SFHs in coarse temporal resolutions~\citep[\emph{e.g.}][]{tojeiro2009, leja2018a}.
While simulations provide another means for understanding SFHs,
they are also subject to their specific time and mass resolutions that
suppress the variability of their star formation, especially in
analytic models, semi-analytic models, and large-volume cosmological
hydrodynamic simulations~\citep[][see also Figure~\ref{fig:illsfh}]{sparre2017}. 

Empirical models, through their flexibility, provide an effective 
method for examining the connection between SFH and host halo assembly 
history. A number of empirical models relate SFHs of galaxies linearly 
to their host halo mass accretion rates and successfully reproduce a 
number of observations~\citep{taghizadeh-popp2015, becker2015, rodriguez-puebla2016a, mitra2017, cohn2017, moster2017}. 
Such models make the strong assumption that SFH of galaxies are perfectly 
correlated to halo accretion history. Recently by analyzing the observed 
correlation between the SFRs and large-scale environment of SF galaxies, 
\cite{tinker2018b} found the first observational evidence for this 
correlation, but with a correlation coefficient of $r \sim 0.63$.
These models, therefore, ignore variation in star formation independent 
from halo accretion, which may come from physical processes in galaxies. 
More recently, the empirical model of \cite{behroozi2019} correlate SFH
with halo assembly while also incorporating star formation variability in
the SFH. For halos at a given $v_{\rm M_{peak}}$ (the maximum circular
velocity of the halo at the redshift of max halo mass) and $z$, they
assign higher SFRs to halos with higher values of $\Delta v_{\rm max}$
(logarithmic growth in the maximum circular velocity of the halo over past dynamical time)
allowing for random scatter in the assignment. Through this random scatter,
which is further separated into contributions from shorter and longer timescales,
they incorporate star formation variability. Explicitly examining and constraining
the timescale of star formation variability, however, is difficult with such a
parameterization. Such constraints can shed light on physical processes 
involved in galaxy star formation and constrain galaxy feedback 
models~\citep{sparre2015}. For instance, it can be used to differentiate 
between physical processes such as galactic feedback interacting with the 
circumgalactic medium, which would cause longer timescale variations, or 
internal processes affecting the cold gas in the galaxy, which would cause 
$\sim 100$ Myr variations. Using the Feedback In Realistic Environments 
(FIRE) high resolution cosmological simulations, \cite{hopkins2014} find 
that explicit and resolved feedback increases time variability in SFRs.
Also using FIRE, \cite{sparre2017} find that varying the strength of
Type II supernova feedback can change the burstiness of SFHs.
\cite{governato2015} find that HI shielding from UV radiation and
early feedback from young stars would also produce small scale star
formation variability.

In this paper, we construct empirical models to investigate the timescale
of star formation variability and the connection between SFH and host halo
accretion history of SF central galaxies. Central galaxies constitute the 
majority of massive galaxies ($M_*>10^{9.5}M_\sun$) at $z\sim0$~\citep{wetzel2013} 
and their SFHs are not influenced by environmentally-driven external 
mechanisms that impact SFHs of satellites such as ram pressure 
stripping~\citep{gunn1972,bekki2009}, strangulation~\citep{larson1980, peng2015}, 
or harassment~\citep{moore1998}. Using a similar approach as~\cite{wetzel2013} and~\cite{hahn2017b}, 
we present models that combine a cosmological $N$-body simulation with observed 
evolutionary trends of the SFS. They statistically 
track the star formation, stellar mass, and host halo assembly histories of 
SF central galaxies from $z\sim1$ to $0$. After fitting our models to reproduce 
the properties of observed SF central galaxies, we compare the predicted 
scatter in the SHMR at $M_h=10^{12}M_\odot$ (hereafter $\sigtwe$) 
of our models to constraints from observations and modern galaxy formation 
models. This comparison allows us to constrain the timescale of star formation 
variability and the correlation between SFH and host halo assembly history. 
Through these comparisons, we examine how our models can produce the constant 
$\sigtwe$ over $z=1$ to 0 found in halo model analyses. We also examine how 
our models can reduce $\sigtwe$ from $z=1$ to 0 as found in the EAGLE~\citep{matthee2017} 
and Illustris TNG~\citep{pillepich2018} hydrodynamic simulations. Lastly 
we investigate the impact of varying $\siglogm(z=1)$. In 
Section~\ref{sec:sdss} we describe the $z\approx0$ central galaxy sample used to compare our models that we 
construct from SDSS Data Release 7. Then in Section~\ref{sec:sim}, we 
describe the $N$-body simulation and how we evolve the SFR and stellar 
masses of the SF central galaxies in our model. We compare predictions 
from our model to observations and present the resulting constraints in 
Section~\ref{sec:results}. Finally, we conclude and summarize the results 
in Section~\ref{sec:summary}.

\section{Central Galaxies of SDSS DR7} \label{sec:sdss}
We construct our galaxy sample following the sample selection of \cite{tinker2011}.
We select a volume-limited sample of galaxies at $z \approx 0.04$ with
$M_r - 5 \log(h) < -18$ and complete above $M_* > 10^{9.4} h^{-2}M_\odot$ from
the NYU Value-Added Galaxy Catalog \citep[VAGC;][]{blanton2005} of the
Sloan Digital Sky Survey Data Release 7~\citep[SDSS DR7;][]{abazajian2009}.
The stellar masses of these galaxies are estimated using the
$\mathtt{kcorrect}$ code~\citep{blanton2007} assuming a~\cite{chabrier2003}
initial mass function. For their specific star formation rates (SSFR) we use
measurements from the current release of the MPA-JHU spectral
reductions\footnote{http://wwwmpa.mpa-garching.mpg.de/SDSS/DR7/}~\citep{brinchmann2004}.
Generally, $\mathrm{SSFR} > 10^{-11}\mathrm{yr}^{-1}$ are derived from
$\mathrm{H}\alpha$ emission, $10^{-11} > \mathrm{SSFR} > 10^{-12}\mathrm{yr}^{-1}$
are derived from a combination of emission lines, and $\mathrm{SSFR} < 10^{-12}\mathrm{yr}^{-1}$
are based on $D_n 4000$~\citep[see discussion in][]{wetzel2013}. We emphasize that
$\mathrm{SSFR} < 10^{-12}\mathrm{yr}^{-1}$ should only be considered upper limits
to the actual galaxy SSFR~\citep{salim2007}.

From this galaxy sample, we identify central galaxies using the
\cite{tinker2011} group finder, a halo-based algorithm that uses
the abundance matching ansatz to iteratively assign halo masses to
groups~\citep[see also][]{yang2005}. Every group contains one
central galaxy, which by definition is the most massive, and a group can
contain zero, one, or many satellites. As with any group finder, galaxies are misassigned due to projection
effects and redshift space distortions. Our central galaxy sample has
a purity of ${\sim}90\%$ and completeness of ${\sim}95\%$~\citep{tinker2018}
Moreover, as illustrated in \cite{campbell2015}, the \cite{tinker2011} group
finder robustly identifies red and blue centrals as a function of stellar mass,
which is highly relevant to our analysis.
We present the SFR--$M_*$ relation of the SDSS DR7 central galaxies, described
above, in the left panel of Figure~\ref{fig:groupcat}. The contours of the
relation clearly illustrate the bimodality in the galaxy sample with the
star-forming centrals lying on the so-call ``star-forming sequence'' (SFS).

\section{Model: Simulated Central Galaxies} \label{sec:sim}
We are interested in constructing a model that tracks central galaxies and
their star formation within the hierarchical growth of their host halos. This
requires a cosmological $N$-body simulation that accounts for the complex
dynamical processes that govern the host halos of galaxies. In this paper
we use the high resolution $N$-body simulation from~\cite{wetzel2013} generated
using the \cite{white2002} $\mathtt{TreePM}$ code with flat $\Lambda$CDM cosmology
($\Omega_m =0.274, \Omega_b = 0.0457, h = 0.7, n=0.95$, and $\sigma_8 = 0.8$).
From initial conditions at $z = 150$, generated from second-order Lagrangian
Perturbation Theory, $2048^3$ particles with mass of $1.98 \times 10^8\,M_\odot$ are
evolved in a $250\,h^{-1}\mathrm{Mpc}$ ($357.12~{\rm Mpc}$) box with a Plummer 
equivalent smoothing of $2.5\,h^{-1}{\rm kpc}$~\citep{wetzel2013, wetzel2014}. 
`Host halos' are then identified using the Friends-of-Friends algorithm~\citep[FoF;][]{davis1985} with
linking length of $b{=}0.168$ times the mean inter-partcile spacing,
which links particles with local density $>{\sim}100\times$ the mean matter density.
Within these host halos, \cite{wetzel2013} identifies `subhalos' as overdensities
in phase space through a six-dimensional FoF algorithm~\citep[FoF6D;][]{white2010}.
The host halos and subhalos are then tracked across the simulation outputs
from $z = 10$ to $0$ to build merger trees~\citep{wetzel2009,wetzel2010}.
The most massive subhalos in newly-formed host halos at a given simulation
output are defined as the `central' subhalo. A central subhalo retains its
`central' definition until it falls into a more massive host halo
(FoF halo mass), at which point it becomes a `satellite' subhalo.

Throughout its $45$ snapshot outs, $\mathtt{TreePM}$ simulation tracks
the evolution of subhalos back to $z \sim 10$. We restrict ourselves to $15$
snapshots from $z = 1.08$ to $0.05$, where we have the most statistically
meaningful observations. Furthermore, since we are interested in centrals we only
keep subhalos that are classified as centrals throughout the redshift
range. This criterion removes ``back splash'' or ``ejected'' satellite
galaxies~\citep[\emph{e.g.}][]{mamon2004,wetzel2014} misclassified as
centrals. Next, we describe how we select and initialize the SF central
galaxies in our model from the central subhalos of the $\mathtt{TreePM}$
simulation.

\subsection{Selecting Star-Forming Centrals}  \label{sec:sfcen}
To construct a model that tracks the SFR and stellar mass evolution of
SF central galaxies, we first need to select them from the
central galaxies/subhalos in the $\mathtt{TreePM}$ simulation. Since
we want our model to reproduce observations, our selection is based
on $f^\mathrm{cen}_\mathrm{SFS}(M_*)$, the fraction of central galaxies
within the SFS measured from the SDSS DR7 VAGC (Section~\ref{sec:sdss}).
Below, we describe how we derive $f^\mathrm{cen}_\mathrm{SFS}(M_*)$ and
use it to select SF central galaxies in our model. Afterwards
we describe how we initialize the SFRs and $M_*$ of these galaxies at
$z = 1$.

Often in the literature, an empirical color-color or SFR--$M_*$ cut
that separates the two main modes (red/blue or star-forming/quiescent)
in the distribution is chosen to classify
galaxies~\citep[\emph{e.g.}][]{baldry2006, blanton2009, drory2009, peng2010, moustakas2013, hahn2015}.
The red/quiescent or blue/star-forming fractions derived from this sort
of classification, by construction, depend on the choice of cut and
neglect galaxy subpopulations such as transitioning galaxies~\emph{i.e.}
galaxies in the ``green valley''. Instead, for our $f^\mathrm{cen}_\mathrm{SFS}(M_*)$,
we use the SFS identified from the \cite{hahn2018a} method, which uses Gaussian
Mixture Models (GMM) and the Bayesian Information Criteria to fit the
SFR--$M_*$ relation of a galaxy population and identify its SFS. This
data-driven approach relaxes many of the assumptions and hard cuts that
go into other methods and can be flexibly applied to a wide range of SFRs 
and $M_*$s and for multiple
simulations. The weight of the SFS GMM component from the method provides
an estimate of $f^\mathrm{cen}_\mathrm{SFS}$. In the right panel of
Figure~\ref{fig:groupcat}, we present the SSFR distribution, $p(\log \mathrm{SSFR})$,
of the SDSS DR7 central galaxies within $10.6 < \log M_* < 10.8$ with
the SFS GMM component shaded in blue.
The SFS constitutes $f^\mathrm{cen}_\mathrm{SFS} = 0.21$ of the SDSS
central galaxies in this stellar mass bin. Using the $f^\mathrm{cen}_\mathrm{SFS}$
estimates, we fit $f^\mathrm{cen}_\mathrm{SFS}$ as a linear function of
$\log M_*$ similar to \cite{wetzel2013,hahn2017b}:
\beq \label{eq:f_cen_sfms}
f^\mathrm{cen}_\mathrm{SFS, bestfit}(M_*) = -0.627\,(\log\,M_* - 10.5) + 0.354.
\eeq
We note that this is in good agreement with the $f_\mathrm{Q}^\mathrm{cen}(M_*; z \sim 0)$
fit from \cite{hahn2017b}.

To select the SF centrals from the subhalos, we begin by assigning $M_*$
at $z\sim 0$ to the subhalos by abundance matching to $M_\mathrm{peak}$,
the maximum host halo mass that it ever had as a central subhalo~\citep{conroy2006,vale2006,yang2009,wetzel2012,leja2013,wetzel2013,wetzel2014,hahn2017b}.
Abundance matching, in its simplest form, assumes a one-to-one mapping between subhalo
$M_\mathrm{peak}$ and galaxy stellar mass, $M_*$, that preserves rank
order: $n({>}M_\mathrm{peak}) > n({>}M_*)$. In practice, we apply a $0.2$
dex log-normal scatter in $M_*$ at fixed $M_\mathrm{peak}$ based on the
observed SHMR~\citep[\emph{e.g.}][]{mandelbaum2006a, more2011, velander2014, zu2015, gu2016, lange2018a}.
For $n({>}M_*)$, we use observed stellar mass function (SMF)
from \cite{li2009} at $z\sim 0$, which is based on the same SDSS NYU-VAGC sample as our
group catalog. Then using the abundance matched $M_*$, we randomly select subhalos as
SF based on the probabilities of being on the SFS using Eq.~\ref{eq:f_cen_sfms}.
\cite{tinker2017b,tinker2018} find that quenching is
independent of halo growth rate and therefore we randomly select SF subhalos.
In our model, we assume that once a SF galaxy quenches its star formation,
it remains quiescent.  
Without any quiescent galaxies rejuvenating their star formation, galaxies
on the SFS at $z\sim0$ are also on the SFS at $z > 0$. Under this assumption
the SF centrals we select at $z \sim 0$ are also on the SFS at the initial
redshift of our model: $z \sim 1$.

We next initialize the SF centrals at $z\sim1$ using the observed SFR-$M_*$
relation of the SFS with $M_*$ assigned using abundance matching with a $z\sim1$ SMF
interpolated between the \cite{li2009} SMF and the SMF from \cite{marchesini2009}
at $z = 1.6$. We choose the \cite{marchesini2009} SMF, among others, because it
produces interpolated SMFs that monotonically increase over $z < 1$. As noted
in \cite{hahn2017b}, at $z \approx 1$, the SMF interpolated between the
\cite{li2009} and \cite{marchesini2009} SMFs is consistent with more recent
measurements from \cite{muzzin2013} and \cite{ilbert2013}. We apply a
$\siglogm^{\rm init} = 0.2$ dex log-normal scatter in the abundance matching based on 
observations~\citep[\emph{e.g.}][]{leauthaud2012, tinker2013, patel2015}.
We next assign SFRs based on $z \sim 1$ observations in the literature.
However, observations, not only use galaxy properties derived differently
from the SDSS VAGC but they also find SFS with significant discrepancies
from one another. In a compilation of SFSs from 25 studies in the literature, 
\cite{speagle2014} find that the SFRs of the SFSs at $z\sim1$ vary by more 
than a factor of 2 at $M_* = 10^{10.5}\, M_\odot$, even after their 
calibration~\citep[see Figure 2 of][]{speagle2014}. With little consensus 
on the SFS at $z\sim1$, and consequently its redshift evolution, we 
flexibly parameterize the SFR of the SFS, $\log\mathrm{SFR}_\mathrm{SFS}(M_*, z)$,
with free parameters that characterize the stellar mass dependence of the SFS
below and above $10^{10} M_\sun$ and the redshift dependence ($m^\mathrm{low}_{M_*}$, 
$m^\mathrm{high}_{M_*}$, and $m_z$, respectively):
\beq \label{eq:logsfr_ms}
\logsfr_{\rm SFS}(M_*, z) =  m_{M_*}\,(\log M_* - 10.) + m_z (z - 0.05) - 0.19
\eeq
\begin{equation*}
{\rm where}~m_{M_*} = \begin{cases}
m^\mathrm{low}_{M_*} & \text{for}\,M_* < 10^{10}M_\sun \\
m^\mathrm{high}_{M_*} & \text{for}\,M_* \geq 10^{10}M_\sun.
\end{cases}
\end{equation*}
We assign SFRs to our SF centrals at $z\sim1$ by sampling a log-normal
distribution centered about $\log\,\mathrm{SFR}_\mathrm{SFS}(M_*, z{=}1)$
with a constant scatter of $0.3\,\mathrm{dex}$ from observations~\citep{daddi2007, noeske2007, magdis2012, whitaker2012}.
Later when comparing to observations, we choose conservative priors
for the parameters $m^\mathrm{low}_{M_*}$, $m^\mathrm{high}_{M_*}$ and $m_z$
that encompass the best-fit SFS from~\cite{speagle2014} as well as measurements
from~\cite{moustakas2013} and~\cite{lee2015}. With our SF centrals initalized
at $z \sim 1$, next, we describe how we evolve their SFR and $M_*$.

\begin{figure}
\begin{center}
\includegraphics[width=0.5\textwidth]{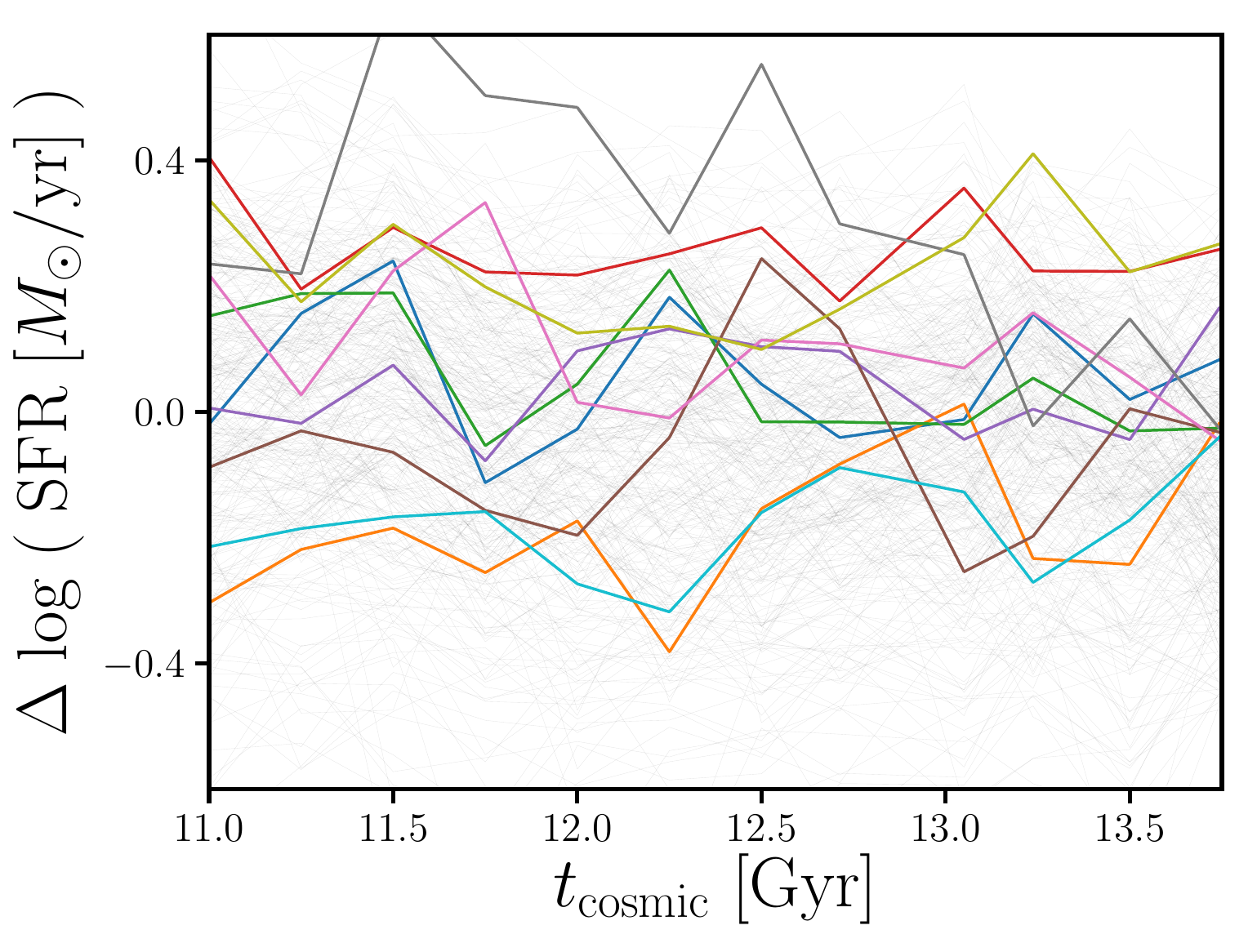}
    \caption{SF galaxies in the Illustris hydrodyanmical simulation have SFHs that evolve 
    along the SFS, with their SFRs stochastically fluctuating about the mean $\logsfr$ 
    of the SFS. We highlight $\Delta \logsfr$, SFR with respect to $\logsfrsfs$ (Eq.~\ref{eq:logsfr_sf}),
    for a handful of galaxies with $10^{10.5}< M_* < 10^{10.6}M_\odot$ at $z\sim0$.
    We calculate $\Delta \logsfr$ with $\logsfrsfs$ identified using the \cite{hahn2018a}
    method, same as in Section~\ref{sec:sfcen}. The implementation of $\mathrm{SFR}$
    variability in the SFHs of SF centrals in our model (Section~\ref{sec:modelevol})
    is motivated by the SFHs of Illustris galaxies above.
    }
\label{fig:illsfh}
\end{center}
\end{figure}

\begin{figure}
\begin{center}
\includegraphics[width=0.85\textwidth]{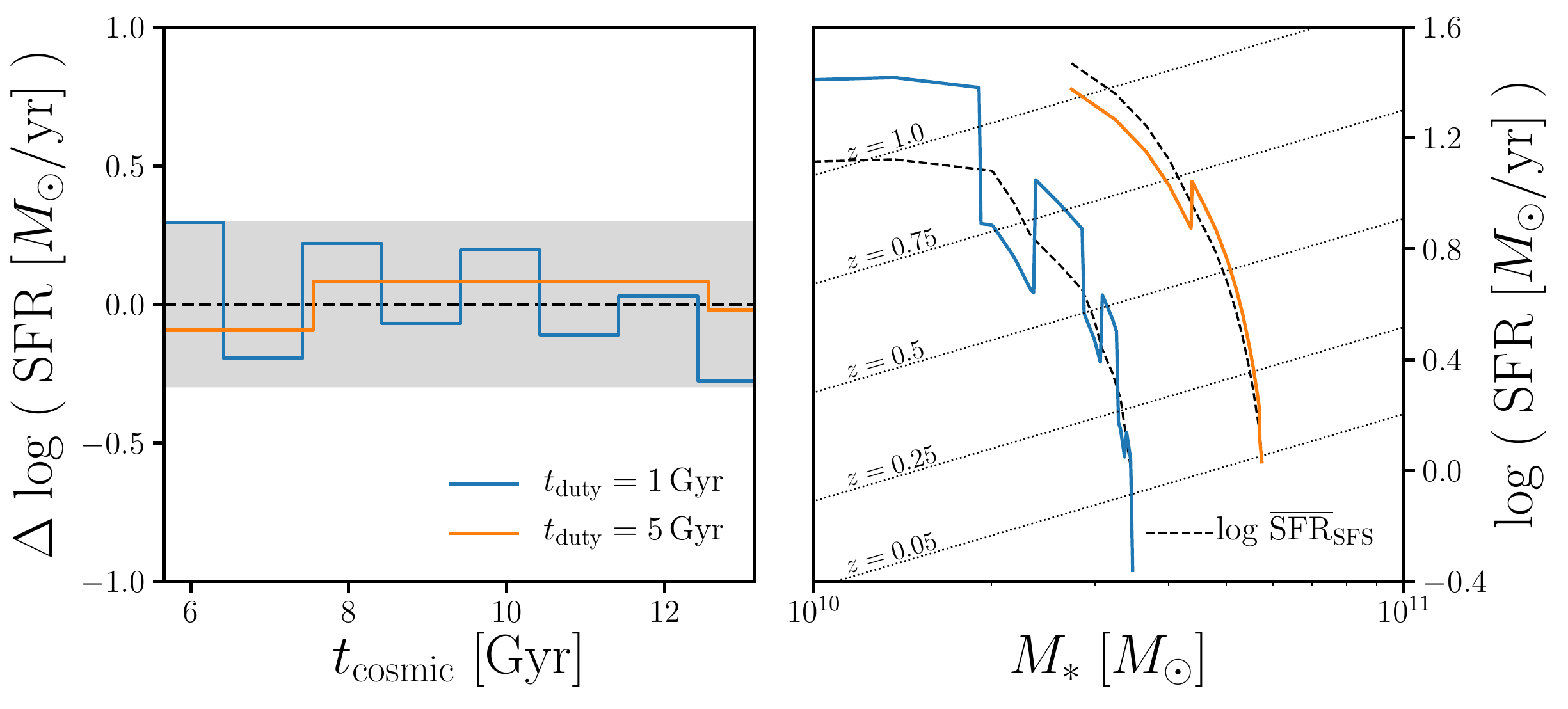}
    \caption{We incorporate star formation variability in our model using a
    ``star formation duty cycle'' where the SFRs of SF centrals fluctuate about
    $\logsfrsfs$ on some timescale $t_\mathrm{duty}$. In our fiducial
    prescription, we randomly sample $\Delta \logsfr$ from a log-normal
    distribution with $0.3$ dex scatter at each duty cycle timestep. We illustrate
    $\Delta\logsfr_i(t)$ of two SF centrals with star formation duty cycles
    on $t_\mathrm{duty} = 1$ (blue) and $5$ Gyr (orange) timescales
    in the left panel. $\Delta \logsfr(t)$ determines the SFH and hence
    the $M_*$ growth of the SF central galaxies (Eq.~\ref{eq:integ_mass}).
    On the right, we illustrate the $\mathrm{SFR}$ and $M_*$ evolutions
    of the corresponding SF centrals. For reference, we include
    $\logsfrsfs(M_{*,i}(t), t)$ that the galaxies' SFR and $M_*$ evolve along
    (black solid). We also include $\logsfrsfs(M_*)$ at various redshifts
    between $z = 1$ to $0.05$ (dotted  lines). \emph{The SF centrals in 
    our model evolve their SFRs and $M_*$ along the SFS with their SFRs 
    fluctuate about $\logsfrsfs$}.} \label{fig:sfh_model}
\end{center}
\end{figure}

\subsection{Evolving along the Star Formation Sequence} \label{sec:modelevol}
The tight correlation between the SFRs and $M_*$ of SF galaxies on 
the SFS has been observed spanning over four orders of magnitude
in stellar mass, with a roughly constant scatter of ${\sim}0.3$ dex, and out
to $z > 2$
(\emph{e.g.}~\citealt{noeske2007,daddi2007,elbaz2007,salim2007,santini2009,karim2011,whitaker2012,moustakas2013,lee2015}; see also references in \citealt{speagle2014}).
This correlation is also predicted by modern galaxy formation models~\citep[][see
\citealt{hahn2018a} and references therein]{somerville2015}. The SFS
naturally presents itself as an anchoring relationship to characterize
the star formation and $M_*$ growth histories of SF galaxies throughout $z < 1$. 
\emph{We, therefore, characterize the SFH of each SF central with respect to 
the $\logsfr$ of the SFS} (Eq.~\ref{eq:logsfr_ms}):
\beq \label{eq:logsfr_sf}
\logsfr_i(M_*, t) = \logsfrsfs(M_*, t) + \Delta\logsfr_i(t).
\eeq
Since SFH determines the $M_*$ growth of galaxies, in this prescription,
$\Delta \logsfr_i(t)$ dictates the SFH and $M_*$ evolution of SF centrals.

One simple prescription for $\Delta \logsfr(t)$ would be to keep $\Delta \logsfr$
fixed throughout $z < 1$ to the offsets from the $\logsfrsfs$ in the
initial SFRs of our SF centrals at $z\sim1$, similar to simple analytic
models such as \cite{mitra2015}. Galaxies with higher than average
initial SFRs continue evolving above the average SFS, while SF centrals
with lower than average initial SFRs continue evolving below the average
SFS. In addition to not being able to reproduce observations, which we
later demonstrate, we also do not find such SFHs in SF galaxies of
hydrodynamic simulations such as Illustris~\citep{vogelsberger2014,genel2014}.
In Figure~\ref{fig:illsfh}, we plot $\Delta \logsfr_i$ of SF 
galaxies in the Illustris simulation as a function of cosmic time. These
galaxies have $10^{10.5} < M_* < 10^{10.6}M_\odot$ at $z=0$.
At each simulation output, we calculate $\Delta \logsfr_i$ using Eq.~\ref{eq:logsfr_sf}
with $\logsfrsfs$ derived from the SFS identified in the simulation
using the \cite{hahn2018a} method, same as in Section~\ref{sec:sfcen}. As the
highlighted $\Delta \logsfr_i$ illustrate, SF galaxies in Illustris evolve
along the SFS with their SFRs fluctuating about $\logsfrsfs$.

Motivated by the SFHs of Illustris SF galaxies, we introduce variability
to the SFHs of our SF centrals in the form of a ``\emph{star formation duty cycle}''--- 
\emph{i.e.} we set the SFRs of SF centrals to fluctulate about 
the SFR of the SFS on some timescale $t_{\rm duty}$.
Within the SFH of Eq.~\ref{eq:logsfr_sf}, we parameterize $\Delta \logsfr_i$ to
fluctuate about the $\logsfrsfs$ on timescale, $t_\mathrm{duty}$, with
amplitude sampled from a log-normal distribution with $0.3\,\mathrm{dex}$
scatter. For our fiducial star formation duty cycle prescription, we randomly
sample $\Delta \logsfr_i$ from a log-normal distribution with $0.3\,\mathrm{dex}$ scatter.
We illustrate $\Delta \logsfr_i(t)$ of SF centrals with our star formation
duty cycle prescription on $t_\mathrm{duty}=1$ Gyr (blue) and $5$ Gyr (orange)
timescales in the left panel of Figure~\ref{fig:sfh_model}.
The shaded region represents the observed $0.3\,\mathrm{dex}$ scatter of
$\logsfr$ in the SFS. By construction, this $\Delta \logsfr$ prescription
reproduces the observed log-normal SFR distribution of the SFS at any point in 
the model. Although, this simplified prescription does not reflect the individual 
SFHs of SF centrals, we seek to statistically capture the stochasticity from 
gas accretion, star-bursts, and feedback mechanisms for the entire SF population.
Measuring $t_\mathrm{duty}$ in the duty cycle parameterization provides us with 
an estimate of the timescale of such star formation variabilities and thus provide 
a useful constraint on the physics of galaxy formation.

Using our fiducial SFH prescription, we evolve both the SFR and $M_*$
of our SF centrals along the SFS. Based on Eq.~\ref{eq:logsfr_sf},
the SFRs of our SF centrals are functions of $M_*$, while $M_*$
is the integral of the SFR over time:
\beq \label{eq:integ_mass}
M_*(t) = f_\mathrm{retain} \int\limits_{t_0}^{t} \mathrm{SFR(M_*, t')}\,\mathrm{d}t' + M_0.
\eeq
$t_0$ and $M_0$ are the initial cosmic time and stellar mass at $z \sim 1$,
respectively. $f_\mathrm{retain}$ here is the fraction of stellar mass
that is retained after supernovae and stellar winds; we use
$f_\mathrm{retain} = 0.6$~\citep{wetzel2013} 
and assume instantaneous recycling such that $f_{\rm retain}$ is applied at all times.
We can now evolve the SFR and
$M_*$ of our SF centrals until the final $z=0.05$ snapshot by
solving the differential equation of Eqs.~\ref{eq:logsfr_sf} and~\ref{eq:integ_mass}.
On the right panel of Figure~\ref{fig:sfh_model}, we present the
$\mathrm{SFR}$ and $M_*$ evolutions of two SF centrals with
$t_\mathrm{duty}=1$ (blue) and $5$ Gyr (orange), same
as the left panel. For reference, we include the mean $\logsfr$ of the SFS
that the galaxies' SFR and $M_*$ evolve along, $\logsfrsfs(M_{*,i}(t), t)$
(black solid). We also include $\logsfrsfs(M_*)$ (dotted lines) at various
redshifts between $z = 1$ to $0.05$. Based on the SFH prescription in our
model, SF centrals evolve their SFRs and $M_*$ along the SFS
with their SFRs fluctuate about $\logsfrsfs$.

\begin{figure}
\begin{center}
\includegraphics[width=0.5\textwidth]{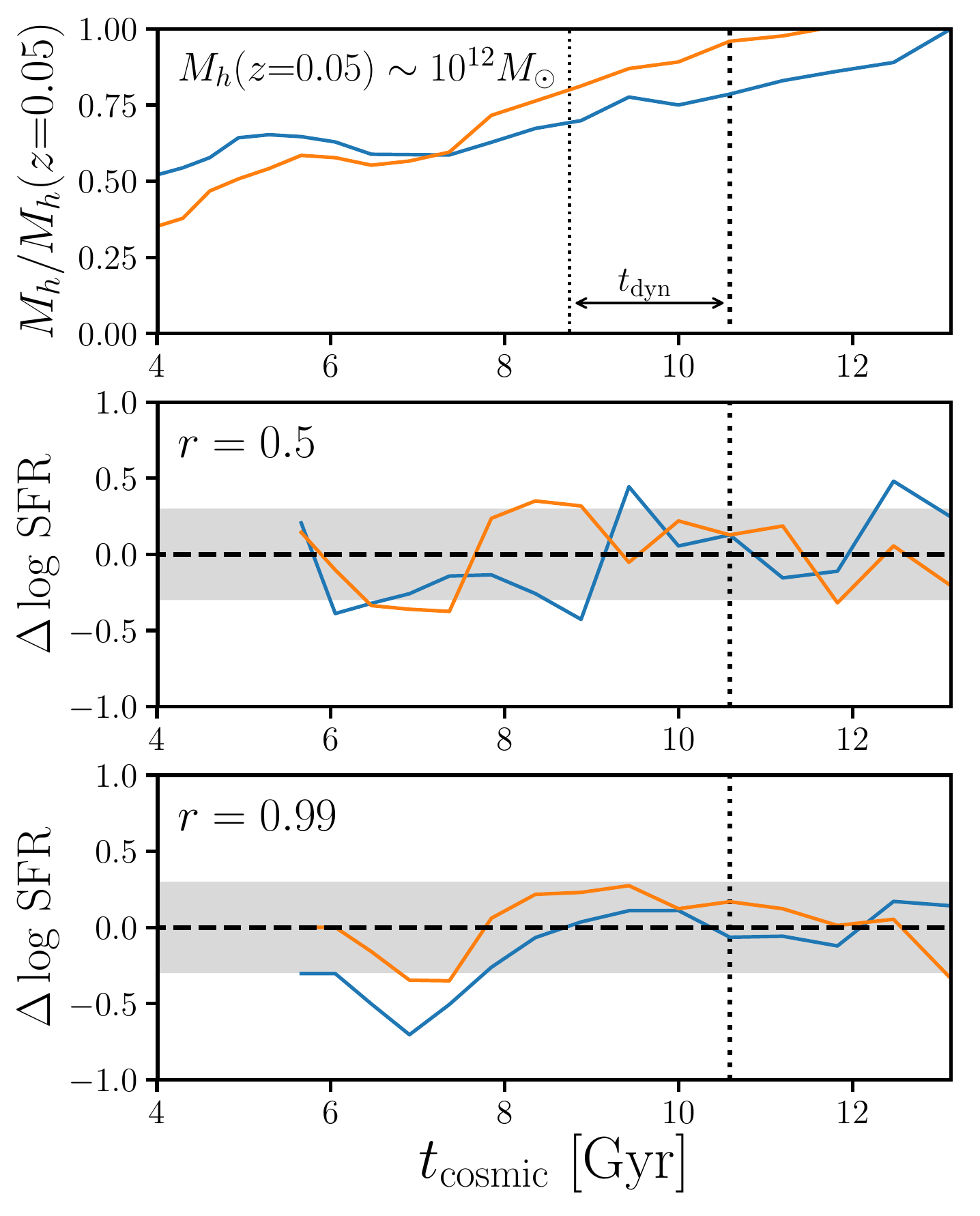}
\caption{
    We incorporate galaxy assembly bias into the SF centrals of our model by
    correlating their host halo accretion history to $\Delta\log{\rm SFR}(t)$, 
    the SFH with respect to the SFS, with correlation coefficient $r$. 
    We plot the relative halo accretion history, $M_h(t)/M_h(z{=}0.05)$
    for two randomly chosen SF centrals with $M_h(z{=}0.05)\sim10^{12}M_\odot$,
    in the top panel. In the two panels below, we present $\Delta\log\,\mathrm{SFR}$,
    of these galaxies for our model with $r=0.5$ and $0.99$ (middle and bottom). 
    The shaded region in these panels mark the $0.3$ dex 1-$\sigma$ width of the 
    log-normal SFS. At some $t$ (dotted),
    $\Delta\log\,\mathrm{SFR}(t)$ is correlated with halo accretion over the
    period $t - t_\mathrm{dyn}$ to $t_\mathrm{dyn}$ labeled in top panel. The
    SFHs illustrate how $\Delta\log\,\mathrm{SFR}(t)$ correlates with
    $\Delta M_h = M_h(t) - M_h(t-t_\mathrm{dyn})$ and how $\Delta\logsfr(t)$
    correlates more strongly with $\Delta M_h(t)$ with higher $r$.}
\label{fig:mhacc_dsfr}
\end{center}
\end{figure}
\subsection{Correlating SFR with Halo Growth}
In our fiducial SFH prescription, we sample $\Delta \logsfr_i$ randomly from a
log-normal distribution with $0.3$ dex scatter. There is, however, growing evidence 
that star formation in galaxies correlate with their host halo accretion 
histories~\citep[\emph{e.g.}][]{lim2016, tojeiro2017, tinker2018b}.
In this section, we introduce \emph{assembly bias} into the SFH prescription of our 
model. Assembly bias, most commonly in the literature, refers to the dependence of the
spatial distribution of dark matter halos on halo properties besides
mass~\citep{gao2005,wechsler2006,gao2007,wetzel2007,li2008,sunayama2016}.
At low halo mass, older and more concentrated halos form in high density environments.
While at high halo mass, the effect is the opposite --- younger, less concentrated
halos form in high-density regions. However, both
simulations~\citep{croton2007, artale2018, zehavi2018} as well as
observations~\citep{yang2006,wang2008,tinker2011,wang2013,lacerna2014,calderon2018,tinker2018},
find that this assembly bias propagates beyond spatial clustering and correlates
with certain galaxy properties such as formation histories and star formation
properties, an effect more specifically referred to as {\em galaxy} assembly bias.
In our model, we incorporate galaxy assembly bias by correlating the SFHs
of our SF central galaxies and their host halo accretion histories
with a correlation coefficient $r$.

We correlate $\Delta\log\,\mathrm{SFR}$ (Eq.~\ref{eq:logsfr_sf}) to the halo 
mass accretion over dynamical time, 
which we define as $t_{\rm dyn} = (\frac{4}{3}\pi G 200\rho_m(t))^{-\frac{1}{2}}$.
At every $t_\mathrm{duty}$ timestep, $t$, $\Delta\logsfr(t)$ is assigned based on 
$\Delta M_h(t) = M_h(t) - M_h(t - t_\mathrm{dyn})$ in $M_\mathrm{max}$ bins
with a correlation coefficient $r$, an additional parameter to our model. 
This prescription for correlating $\Delta\log\mathrm{SFR}$ to $\Delta M_h$ is
similar to other empirical models that also correlate $\Delta\log\mathrm{SFR}$
to $\Delta M_h$ over $t_\mathrm{dyn}$~\citep{rodriguez-puebla2016a, behroozi2019}.
In \cite{rodriguez-puebla2016a}, however, they assume perfect ($r=1$) correlation
between SFH and halo accretion. In the \cite{behroozi2019} {\sc UniverseMachine}
(hereafter UM), $r$ is free parameter and their SFH includes SF variability,
similar to our model. As we mention in the introduction, their prescription,
however, does not not focus on a star formation variation on specific timescales,
which our models do through the star formation duty cycle.

In Figure~\ref{fig:mhacc_dsfr}, we illustrate our prescription for galaxy
assembly bias in our model. We plot the relative halo accretion histories
$M_h(t)/M_h(z{=}0.05)$ of two arbitrarily chosen SF centrals with
$M_h(z{=}0.05)\sim10^{12}M_\odot$ in the top panel (orange and blue). Below, we plot
$\Delta\logsfr$, SFH with respect to the SFS, of these galaxies for our model with
correlation coefficients $r=0.5$ and $0.99$ (middle and bottom). We choose a
random $\mathtt{TreePM}$ snapshot, $t$ (dotted), and label the period
[$t$, $t - t_\mathrm{dyn}$]. Halo accretion over this period,
$\Delta M_h = M_h(t) - M_h(t-t_\mathrm{dyn})$, correlates with $\Delta\logsfr(t)$.
The SFHs in the middle and bottom panels illustrate this correlation and how
$\Delta\logsfr(t)$ correlates more strongly with $\Delta M_h(t)$ for our model
with higher $r$.

\subsection{SHMR scatter at $z=1$} \label{sec:siglogm_init}
So far in both our fiducial and galaxy assembly bias added models above, 
we assume that the log-normal scatter in $M_*$ at fixed $M_h$ at $z\sim1$:
$\siglogm^{\rm init} = 0.2$ dex. This initial condition determines
the initial abundance matching $M_*$ at $z\sim1$ that initializes our models and is motivated
by constraints on the observed SHMR~\citep[\emph{e.g.}][]{leauthaud2012, tinker2013, patel2015}. 
However, these constraints are derived using halo models in which the scatter 
is a constant, independent of $M_h$. For these models, the constraining power 
mainly come from massive halos. Hence, $0.2$ dex does not accurately reflect 
the SHMR scatter at $z\sim1$ for less massive halos 
($M_h \lesssim 10^{12}M_\odot$). Later in this paper, we focus on 
$\sigtwe(z{=}0)$ predicted by our models. We 
therefore examine the impact of varying $\siglogm^{\rm init}$ on 
$\sigtwe$ using models with $\siglogm^{\rm init} = 0.35$ and 0.45 dex. 
Our choice of $\siglogm^{\rm init}$ is
based on the $z=1$ SHMR in the Illustris TNG which has $\sigma_{\log\,M_*}(z\sim1)$
spanning $0.45$ to $0.3$ dex for $M_h = 10^{11.5}$ to $10^{12}M_\odot$.

All of the models we present in this section track the SFRs and $M_*$ of SF
central galaxies. We can compare these properties of our model galaxies
to observed galaxy population statistics (quiescent fraction and SMF) to
constrain the free parameters. Our models, run with these inferred parameters, 
can then be compared to observations of the galaxy-halo connection such as the 
SHMR. In the following section, we present this comparison and the constraints 
we derive on the role and timescale of star formation variability in SF central 
galaxies.

\begin{figure}
\begin{center}
\includegraphics[width=\textwidth]{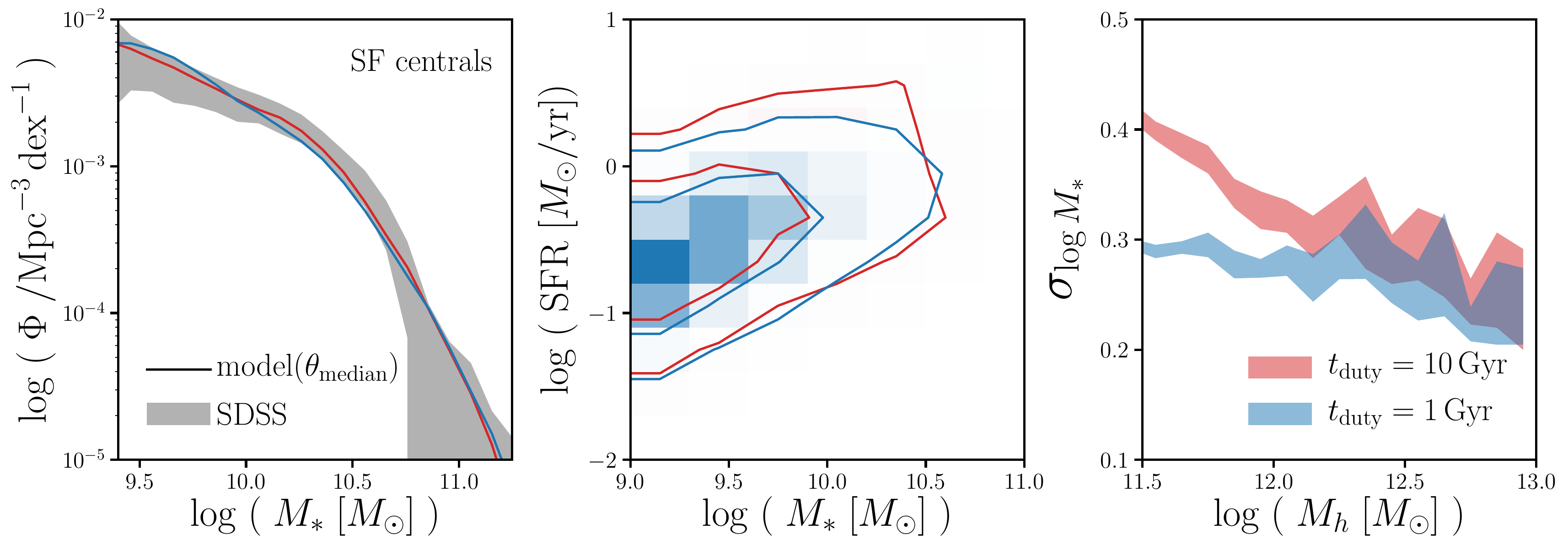}
    \caption{Our models with different star formation duty cycle timescales
    ($t_{\rm duty} = 1$ and 5 Gyr; blue and red) run with median values of 
    their ABC posterior distribution produce SMFs and SFSs consistent with 
    observations (left and middle). \emph{They however predict significantly 
    different scatter in $\log\,M_*$ at fixed $\log\,M_\mathrm{halo}$ for SF 
    centrals --- $\siglogm$ (right)}. By comparing the scatter in SHMR of our 
    models to observational constraints on the SHMR, we can constrain the 
    timescale of the star formation duty cycle and thereby the SFHs of star
    forming galaxies.
    }
\label{fig:abc_demo}
\end{center}
\end{figure}

\section{Results and Discussion} \label{sec:results}
Our models take $\mathtt{TreePM}$ central subhalos and tracks their SFR
and $M_*$ evolution using flexible parameterizations of the SFS and SFHs
that incorporate variability through a star formation duty cycle.
At $z = 0.05$, the final timestep, our models predict SFR and $M_*$ of SF
centrals, along with their host halo properties. We now use these predicted 
properties to compare our model to observations and constrain its free
parameters --- the SFS parameters of Eq.~\ref{eq:logsfr_ms}. Since we focus
on SF centrals, for our observations we use the SMF of SF centrals in SDSS, 
which we estimate as \beq \label{eq:smf_sf_cen}
\Phi^\mathrm{SDSS}_\mathrm{SF,cen} = f^\mathrm{cen}_\mathrm{SFS} \times f_\mathrm{cen} \times \Phi^\mathrm{Li\&White(2009)}.
\eeq
$f^\mathrm{cen}_\mathrm{SFS}$ is the fraction of central galaxies on the
SFS, which we fit in Eq.~\ref{eq:f_cen_sfms}. $f_\mathrm{cen}$ is the
central galaxy fraction from \cite{wetzel2013} and $\Phi^\mathrm{Li\&White(2009)}$
is the SMF of the SDSS from \cite{li2009}. If our models reproduce the
observed $\Phi^\mathrm{SDSS}_\mathrm{SF,cen}$, by construction they reproduce
the observed quiescent fraction.

For the comparison between our models and observation, we use the likelihood-free parameter
inference framework of Approximate Bayesian Computation (ABC). ABC has the
advantage over standard approaches to parameter inference in that it does not
require evaluating the likelihood. For observables with likelihoods that are
difficult or intractable, incorrect assumptions in the likelihood can significantly
bias the posterior distributions~\citep[\emph{e.g.}][]{hahn2018}. Instead,
ABC relies only on a simulation of the observed data and a distance metric to
quantify the ``closeness'' between the observed data and simulation. Many variations
of ABC has been used in astronomy and
cosmology~\citep[\emph{e.g.}][]{cameron2012,weyant2013,ishida2015,alsing2018}.
We use ABC in conjunction with the efficient Population Monte Carlo (PMC)
importance sampling as in~\cite{hahn2017b, hahn2017a}. For initial sampling
of our ABC particles, \emph{i.e.} the priors of our free parameters
$m^\mathrm{low}_{M_*}$, $m^\mathrm{high}_{M_*}$, and $m_z$, we use uniform
distributions over the ranges $[0.0, 0.8]$, $[0.0, 0.8]$, and
$[0.5, 2.0]$, respectively. The range of the prior were conservatively chosen
to encompass the best-fit SFS from~\cite{speagle2014} as well as 
measurements from~\cite{moustakas2013} and~\cite{lee2015} at $z \sim 1$.
Finally, for our distance metric we use the following distance between
the SMF of the star-forming centrals in our model to the observed
$\Phi^\mathrm{SDSS}_\mathrm{SF,cen}$:
\beq
\rho_\Phi = \sum\limits_{M} \left( \frac{\Phi^\mathrm{sim} - \Phi^\mathrm{SDSS}_\mathrm{SF,cen}}{\sigma'_\Phi}\right)^2.
\eeq
$\Phi^\mathrm{sim}(M)$ above is the SMF of the SF centrals in our model
and $\sigma'_\Phi(M)$ is the uncertainty of $\Phi^\mathrm{SDSS}_\mathrm{SF,cen}$,
which we derive by scaling the~\cite{li2009} uncertainty of $\Phi^\mathrm{SDSS}$
derived from mock catalogs. 
For the rest of our ABC-PMC implementation, we strictly follow the implementation
of~\cite{hahn2017a} and \cite{hahn2017b}. We refer reader to those papers for
further details.

\begin{figure}
\begin{center}
\includegraphics[width=0.75\textwidth]{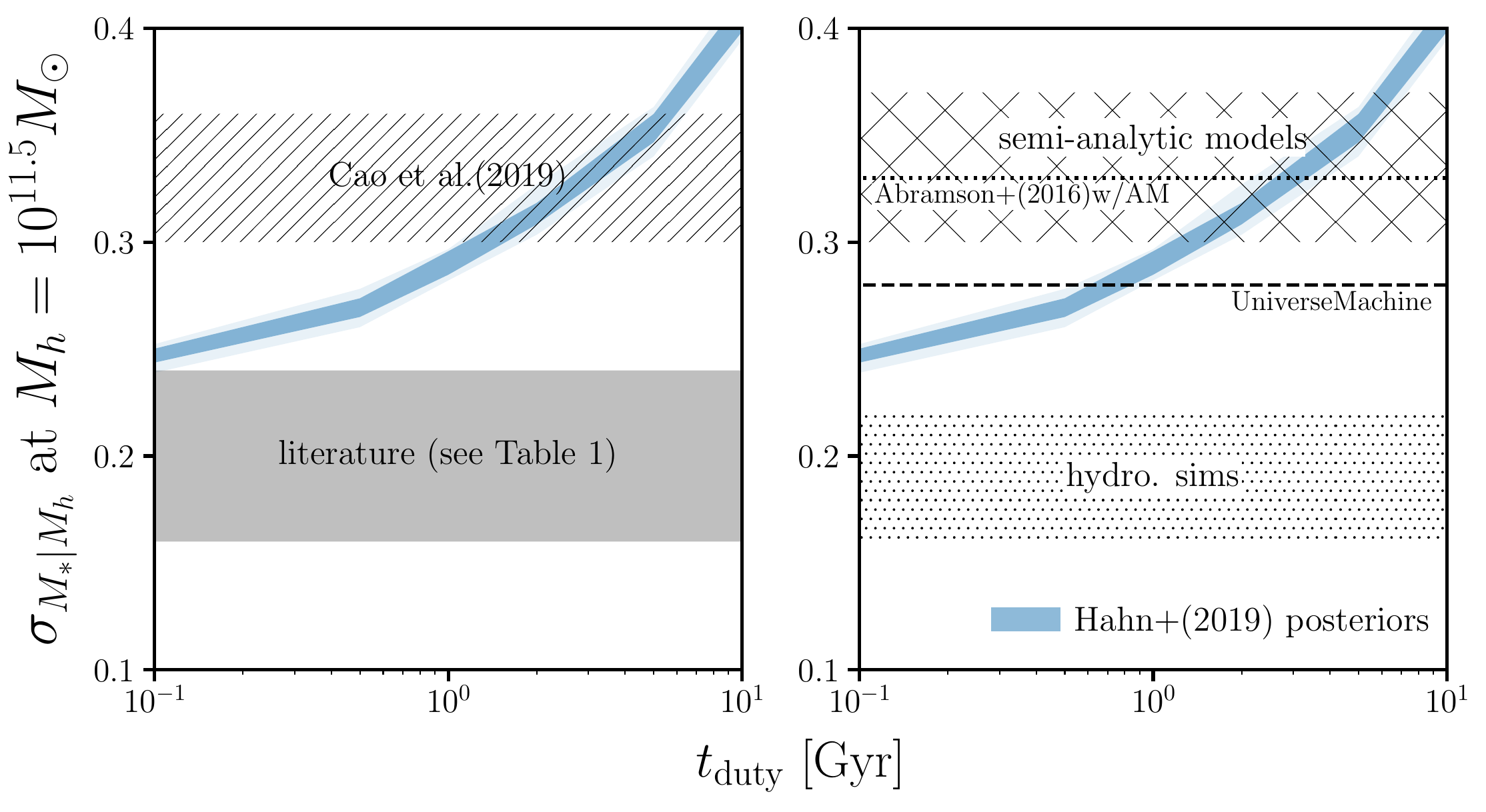}
    \caption{With shorter star formation duty cycle timescales, $t_\mathrm{duty}$,
    our fiducial model predicts smaller scatter in $\log M_*$ at $M_h = 10^{12} M_\odot$ --- $\sigtwe$ (blue). 
    The dark and light blue shaded regions represent the $68\%$ and $95\%$ confidence
    intervals of the predicted scatter from the ABC posteriors of our
    model with $t_\mathrm{duty} = 0.1 - 10$ Gyr. For $t_\mathrm{duty} = 10$
    to $0.1$ Gyr, $\sigtwe$ ranges from $0.32\substack{+0.02\\ -0.02}$ to
    $0.26\substack{+0.01\\-0.01}$. In the left panel, we include for comparison
    observational $\siglogm$ constraints from \cite{leauthaud2012, zu2015, tinker2017, lange2018a}
    within the shaded region and Cao et al. (in preparation) in the diagonally 
    hatched region (see Section~\ref{sec:sfdutycycle} and~\ref{tab:lit}).
    In the right panel, we include compiled predictions from hydrodynamic simulations
    (dotted region; EAGLE, Massive Black II, Illustris TNG), semi-analytic models (hatched), and
    the~\cite{behroozi2019} {\sc UM} empirical model. We also include 
    $\sigma_{M_*|M_h = 10^{12.4} M_\odot}$ from a simple empirical model with \cite{abramson2016} 
    SFHs assigned to halos via abundance matching (dotted).
    \emph{A shorter $\tduty$ produces significantly tighter scatter in the SHMR.
    Halo model observations constraints and predictions from hydrodynamic simulations
    favor a star formation variability on $\tduty \lesssim 0.1$ Gyr for our fiducial
    model.}
    }
\label{fig:sigMstar_duty}
\end{center}
\end{figure}

\begin{table}
    \caption{$\siglogm$ constraints in the literature} 
    \begin{center}
        \begin{tabular}{c p{6cm} p{2cm} c} \toprule
            & observations & method & $\siglogm$  \\[3pt] \hline\hline
            \cite{leauthaud2012}    &COSMOS: SMF, galaxy clustering, galaxy-galaxy lensing& HOD & $0.206\substack{+0.031\\-0.021}$\\\hline 
            \cite{reddick2013}      &SDSS DR7: satellite kinematics, projected galaxy clustering, conditional SMF & abundance matching & $0.21\substack{+0.025\\-0.025}$ \\\hline 
            \cite{zu2015}           &SDSS DR7: galaxy clustering, galaxy-galaxy lensing& HOD & $0.22\substack{+0.02\\-0.02}$ \\\hline 
            \cite{tinker2017}       &BOSS: projected galaxy clustering & abundance matching & $0.18\substack{+0.01\\-0.02}$ \\\hline
            \cite{lange2018a}$^{*}$ &SDSS DR7: conditional luminosity function, radial profile of satellite galaxies& HOD & $0.23\substack{+0.018\\-0.018}$ \\\hline
            Cao et al. in prep.     &\multicolumn{2}{c}{SDSS DR7: kurotsis of line-of-sight pairwise velocity distribution} & $0.33\substack{+0.03\\-0.03}$ \\ [3pt]
    \hline
\end{tabular} \label{tab:lit}
\end{center}
    $^*$\cite{lange2018a} constrain $\sigma_{L|M_h}$ instead of $\sigma_{M_*|M_h}$. 
\end{table}
\subsection{The Fiducial Model} \label{sec:sfdutycycle}
We present the SMFs (left), SFSs (center), and $\siglogm(M_h)$ (right) of our 
fiducial model run using SFHs with 
$t_\mathrm{duty}{=}10$ (red) and $1\,\mathrm{Gyr}$ (blue) duty cycle timescales 
in Figure~\ref{fig:abc_demo}. 
For each $\tduty$, we evaluate our fiducial model using the median of the posterior parameter 
distributions derived from ABC. For both $\tduty$, our model successfully produces 
SMFs, $\Phi^{\rm SDSS}_{\rm SF,cen}$, and SFSs consistent with observations (left 
and center panels). Despite reproducing observations, the fiducial model with different 
$t_{\rm duty}$ predict significantly different $\siglogm$, 
particularly below $M_h < 10^{12.5}M_\odot$. We further illustrate the sensitivity 
of $\siglogm$ predicted by our model to $\tduty$ in Figure~\ref{fig:sigMstar_duty}, 
where we present $\siglogm$ at fixed $M_h = 10^{12} M_\odot$ for our 
model with $t_{\rm duty} = 0.1 - 10$ Gyr. $\siglogm$ at $t_{\rm duty}$
is the prediction from our model run using parameters from the corresponding ABC 
posterior distributions. The dark and light blue shaded regions represents the 
$68\%$ and $95\%$ confidence intervals. For $t_{\rm duty} = 10$ to $0.1$ Gyr,
$\sigtwe$ ranges from $0.32\substack{+0.02\\ -0.02}$ to $0.26\substack{+0.01\\-0.01}$
--- a shorter star formation duty cycle timescale produces significantly
tighter scatter in the SHMR. 
This difference is even larger for $\siglogm$ at $M_h=10^{11.5}M_\odot$: for 
$t_{\rm duty} = 10$ to 0.1 Gyr, $\sigma_{M_*|M_h=10^{11.5}M_\odot} = 0.41\substack{+0.01\\-0.01}$
to $0.25\substack{+0.004\\-0.003}$. We focus on $\sigtwe$ where there are more
robust constraints from observations and galaxy formation models. 
Hence, \emph{the scatter in the SHMR, particular $\sigma_{M_*|M_h < 10^{12}M_\odot}$, can be used to probe the star variability timescale and SFH of SF central galaxies.}

On the left panel of Figure~\ref{fig:sigMstar_duty}, we compare $\sigtwe$
predictions from our fiducial model to observational constraints in the 
literature. These constraints are mainly derived from fitting halo-occupation 
based models to observations of galaxy clustering, SMF, satellite kinematics, 
or galaxy-galaxy weak lensing. We include constraints 
from~\cite{leauthaud2012, reddick2013, zu2015, tinker2017}, and \cite{lange2018a}
in the shaded region as well as \cao in the diagonally hatached region. 
The width of the shaded and hatched regions encompass the 1-$\sigma$ 
uncertainties of the constraints.
\cite{reddick2013}, and \cite{zu2015} fit SDSS DR7 measurements of satellite 
kinematics, projected galaxy clustering and conditional SMF, and galaxy 
clustering and galaxy-galaxy lensing, respectively. \cite{tinker2017} 
similarly fit the projected galaxy clustering of the Baryon Oscillation 
Spectroscopic Survey~\citep{dawson2013}. Meanwhile,~\cite{leauthaud2012}
use COSMOS to fit the SMF, galaxy clustering, and galaxy-galaxy lensing.
Finally, \cao~fit the kurtosis of the line-of-sight pairwise velocity 
dispersion between central galaxies and all neighboring galaxies to constrain 
the scatter in SHMR at low halo masses. We note that
\cite{leauthaud2012, reddick2013, zu2015, tinker2017} measure $\siglogm$ for all central 
galaxies, not only SF. However, \cite{tinker2013} find little ($< 1\sigma$) difference 
in $\siglogm$ between SF and quiescent centrals, so we include these constraints 
in our comparison. We also include the \cite{lange2018a} constraint from fitting 
color-dependent conditional luminosity function and radial profile of satellite 
galaxies of SDSS DR7. This constraint, however, is on the scatter in luminosity, 
$\log L$, not $\log M_*$ at a given $M_h$. We list the constraints from the 
literature in Table~\ref{tab:lit}. 

Overall, observational constraints are more consistent with $\siglogm$ 
predictions of our fiducial model with a short, $< 1$ Gyr, duty cycle timescale. 
However, there is no clear consensus among the observed $\siglogm$ constraints. 
Besides \cao, the $\siglogm$ constraints in the literature are loosely 
consistent with $\sim 0.2$ dex. These constraints, however, are mostly derived 
using halo models that assume $\siglogm$ is a constant, independent of $M_h$. 
The constraining power for these constraints mainly come from high mass halos 
and, thus, do not reflect $\siglogm$ at $M_h=10^{12}M_\odot$.
While \cite{reddick2013} constrain $\siglogm$ for different bins of
$M_h$ over the range $10^{12} - 10^{14} M_\odot$, their constraints mainly come 
from halos with $M_h \ge 10^{13}M_\odot$~\citep{wechsler2018}. The constraint from
\cite{lange2018a} is also derived from a halo model with $M_h$ dependence. However, 
as mentioned above, they constrain $\sigma_{L|M_h}$. In \cite{more2011}, where 
they constrain both $\sigma_{L|M_h}$ and $\siglogm$ from the same sample, 
they find $\siglogm = 0.15\substack{+0.08\\ -0.11} < \sigma_{L|M_h} = 0.21\substack{+0.06\\ -0.04}$
for blue centrals. Translating $\sigma_{L|M_h}$ to $\siglogm$, however, 
is tenuous for different data sets and models. We also note observational constraints 
include significant measurement uncertainties in $M_*$. The intrinsic $\siglogm$ of 
these constraints, \emph{i.e.} the scatter predicted by our model, will be lower. 
If we consider $0.1 - 0.2$ dex uncertainties in $M_*$~\citep{roediger2015}, the
Cao et al. (in preparation) constraint, for instance, will be reduced from
$\siglogm = 0.33$ dex to $0.31 - 0.26$ dex. 

In addition to the observational constraints, we also compare the $\sigtwe$
predicted by our model to predictions from modern galaxy formation models
on the right panel: hydrodynamic simulations (dot filled), semi-analytic models
(SAM; cross hatched), and an empirical model (dashed line). For the large-volume hydrodynamic 
simulations, the dotted region, $\sigtwe = 0.16 - 0.22$ dex, encompasses
predictions from EAGLE~\citep{matthee2017}, Massive Black II~\citep{khandai2015},
and Illustris TNG, as compiled in Figure 8 of \cite{wechsler2018}.
For the SAMs, the cross hatched region, $\sigtwe = 0.3 - 0.37$ dex, includes predictions 
from~\cite{lu2014, somerville2012} and the SAGE\footnote{\url{https://tao.asvo.org.au/tao/}}
model~\citep{croton2016}. We also include the prediction from the
\cite{behroozi2019} UM empirical model. Similar to observations, there is 
little consensus among the $\siglogm$ predictions of the galaxy formation models.
$\siglogm$ from SAMs are consistent with our model with $t_{\rm duty} \gtrsim 5$ Gyr. 
UM, which predicts a lower $\siglogm$, is consistent with $t_{\rm duty} = 1 - 5$ Gyr.
Lastly, large-volume hydrodynamic simulations predict the lowest $\siglogm$ among the models
with $\sim 0.2$ dex, which our fiducial model struggles to reproduce even with 
$t_{\rm duty} = 0.1$ Gyr. We note that while there is yet no consensus among the 
observational $\siglogm$ constraints at $M_h=10^{12} M_\odot$, at higher $M_h$ 
observations are in better agreement and {\em only} hydrodynamic simulations 
predict $\siglogm$ consistent with these observations~\citep{wechsler2018}. Right 
below $M_h=10^{12}M_\odot$, however, hydrodynamic simulations predict significantly 
higher scatter --- $\siglogm(M_h\sim 10^{11.5}) = 0.22 - 0.32$ dex~\citep{wechsler2018}.

Given the little consensus among the actual $\siglogm$ constraints at $z=0$ 
from both observations and simulations, we examine the redshift evolution trend 
of $\siglogm$ from $z=1$ to 0. For our model $\siglogm$ at $z=1$ 
is an input initial condition we
use to determine the initial abundance matching $M_*$ of our model that we set to 0.2 dex, based 
on halo model observations~\citep[\emph{e.g.}][]{leauthaud2012, tinker2013, patel2015}. 
According to our predictions, $\sigtwe$ increases by $0.06 - 0.12$ dex for 
$t_{\rm duty} = 0.1-10$ Gyr. In comparison, halo model observational constraints find 
constant $\siglogm=0.2$ dex evolution. Meanwhile in Illustris TNG, $\sigtwe$ 
decreases over the redshift range from $\sim 0.3$ dex at $z=1$ to $\sim 0.2$ dex 
at $z=0$ (Cao et al. in preparation). Both of these $\siglogm$ evolution 
trends, despite their difference, favor a short duty cycle. However, even with the 
shortest duty cycle, we find an increasing $\sigtwe$ from $z=1$. 
 
\subsection{Log-Normal SFH}
A key element of our models is the SFH prescription for SF central galaxies where
the SFH evolves about the SFS. Contrary to our SFH prescription, \cite{kelson2014},
for example, argue that the SFS is a consequence of central limit theorem
and can be reproduced even if \emph{in situ} stellar mass growth is modeled as
a stochastic process like a random walk. \cite{gladders2013,abramson2015,abramson2016},
similarly argue that $\sim2000$ loosely constrained log-normal SFHs can reproduce
observations such as the SMF at $z \leq 8$ and the SFS at $z \leq 6$. These works,
however, focus on reproducing observations of galaxy properties and do not examine
the galaxy-halo connection such as the SHMR. In order to test whether log-normal
SFHs can also produce realistic SHMRs,
we construct a simple empirical model using  the SFHs, $\mathrm{SFR}(t)$ and
$M_*(t)$, from \cite{abramson2016} and assign them to halos by abundance matching
their $M_*$ to $M_h$ at $z{\sim}1$.
We then restrict the SFHs to those that would be classified as SF based
on a $\log\,\mathrm{SSFR} > -11$ cut. Afterwards we measure $\siglogm$
at the lowest $M_h$ where it can be reliably measured given the \cite{abramson2016}
sample's $M_*{>}10^{10}M_\odot$ limit.
We find that the \cite{abramson2016} based empirical model predicts a scatter
of $\siglogm(M_h=10^{12.4}M_\odot) = 0.33\pm0.04$ (dotted; right panel of Figure~\ref{fig:sigMstar_duty}).
Although the \cite{abramson2016} SFHs can reproduce various galaxy properties,
the empirical model we construct with them struggles to produce $\siglogm$ comparable 
to observational constraints and predictions from UM and hydrodynamic simulations.
It also struggles to keep $\siglogm$ evolution constant or decreasing with redshift. 
The \cite{abramson2016} based empirical model that we explore utilizes a simple
abundance matching scheme. \cite{diemer2017} find that their their log-normal fits
to the SFHs of Illustris galaxies correlate with halo formation histories. Incorporating
such correlations into the abundance matching may reduce $\siglogm$.

In this section we demonstrate that star formation variability in the SFH impacts
$\sigtwe$: star formation variability on shorter timescales significantly 
reduces $\sigtwe$. Given this dependence, $\siglogm$ can conversely be used to constrain
the timescale of star formation variability. Although there is no clear consensus in 
the $\sigtwe$ of observations or simulations, overall they favor our model with short 
variability timescales $\sim0.1$ Gyr. However, we find that star formation variability 
alone is insufficient in producing the tight SHMR scatter and $\siglogm$ redshift evolution 
trend found in halo model observations and hydrodynamic simulations. In the next section, 
we explore how correlation between SFH and halo formation histories impacts $\sigtwe$
using our models with galaxy assembly bias.

\begin{figure}
\begin{center}
\includegraphics[width=0.75\textwidth]{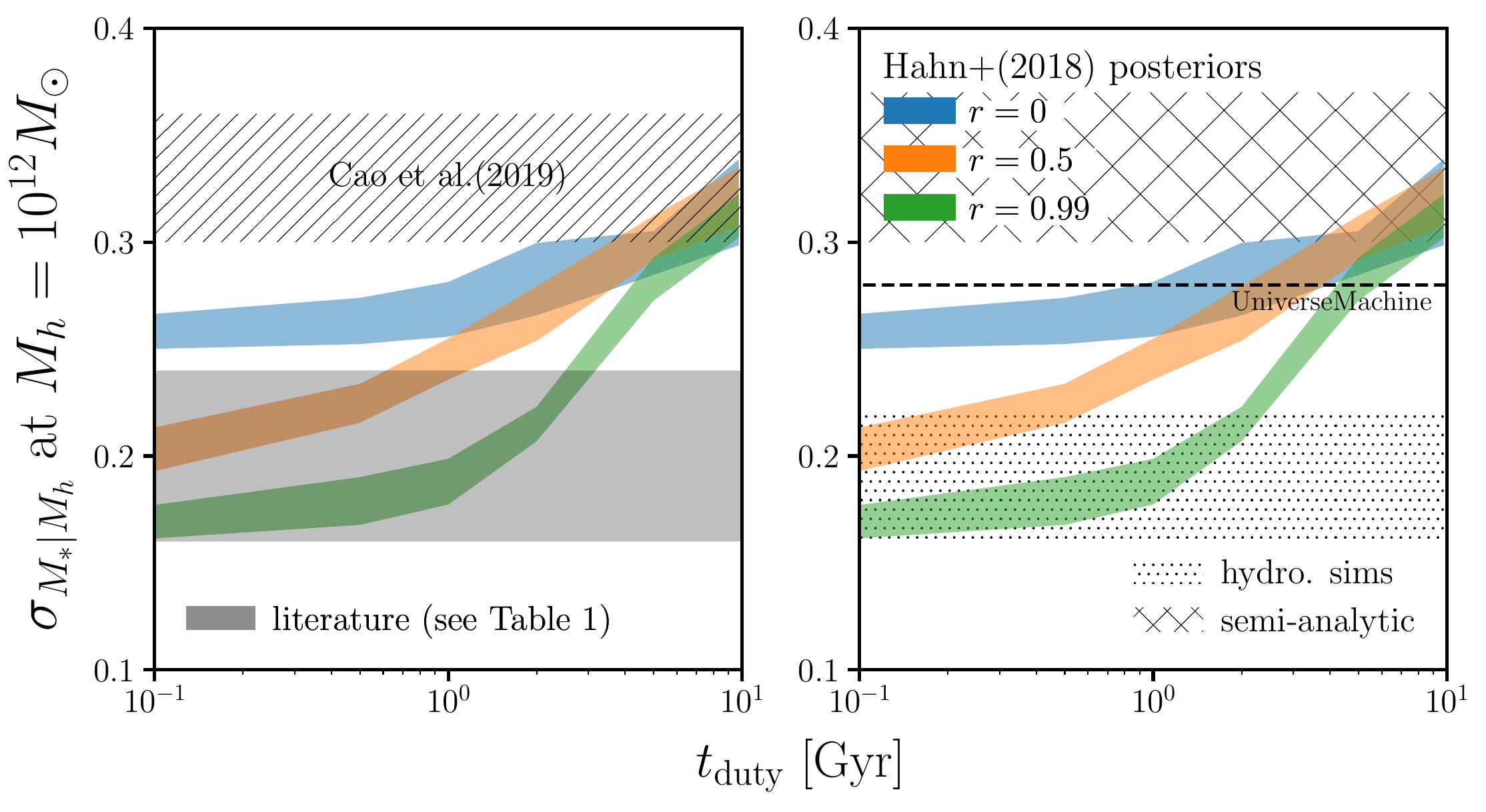}
\caption{
    Using models that correlate SFH with halo assembly history, we find that 
    higher $r$, \emph{i.e.} stronger galaxy assembly bias, significantly reduces the 
    scatter in SHMR for $t_\mathrm{duty} < 5$ Gyr. We plot $\sigtwe$ as a function 
    of the star formation duty cycle timescale, $t_\mathrm{duty}$, for our 
    models with $r = 0$ (no assembly bias; blue), 0.5 (orange), and 0.99 (green).
    We include observational constraints and predictions from galaxy formation models
    in the left and right panels, respectively. With $r > 0.5$, our models can predict 
    $\sigtwe$ more consistent with the tight $\sim0.2$ dex constraint 
    from halo model observations and hydrodynamic simulations. $r > 0$ also reduces the 
    growth in $\sigtwe$ evolution from $z = 1$ to 0.
    }
\label{fig:sigMstar_duty_abias}
\end{center}
\end{figure}

\begin{figure}
\begin{center}
\includegraphics[width=0.45\textwidth]{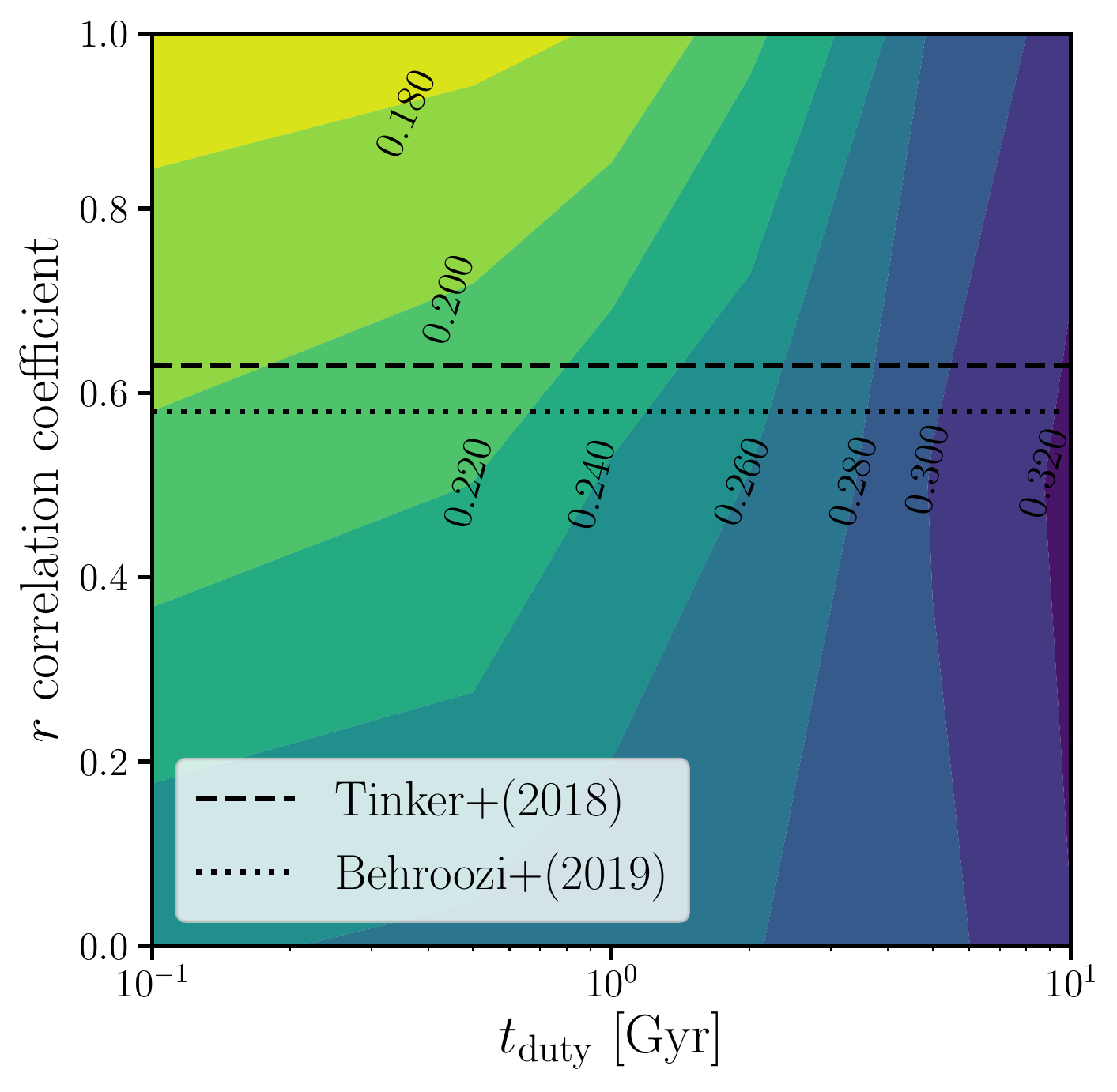}
    \caption{Predicted $\siglogm$ as a function of $\tduty$ and $r$ for our
    models illustrate the degeneracy between the timescale of SF variability and
    the correlation between SFH and halo assembly history. Based on $r$ constraints
    from \cite{tinker2018b} (dashed) and \cite{behroozi2019} (dotted), $\tduty < 0.5$ Gyr
    is necessary to produce $\siglogm \sim 0.2$ dex from observations and
    hydrodynamic simulations. Meanwhile, $\tduty < 5$ Gyr is necessary to produce
    $\siglogm$ from Cao et al. (in preparation), SAMs, and UM.
    }
\label{fig:r_tduty}
\end{center}
\end{figure}

\begin{figure}
\begin{center}
    \includegraphics[width=0.95\textwidth]{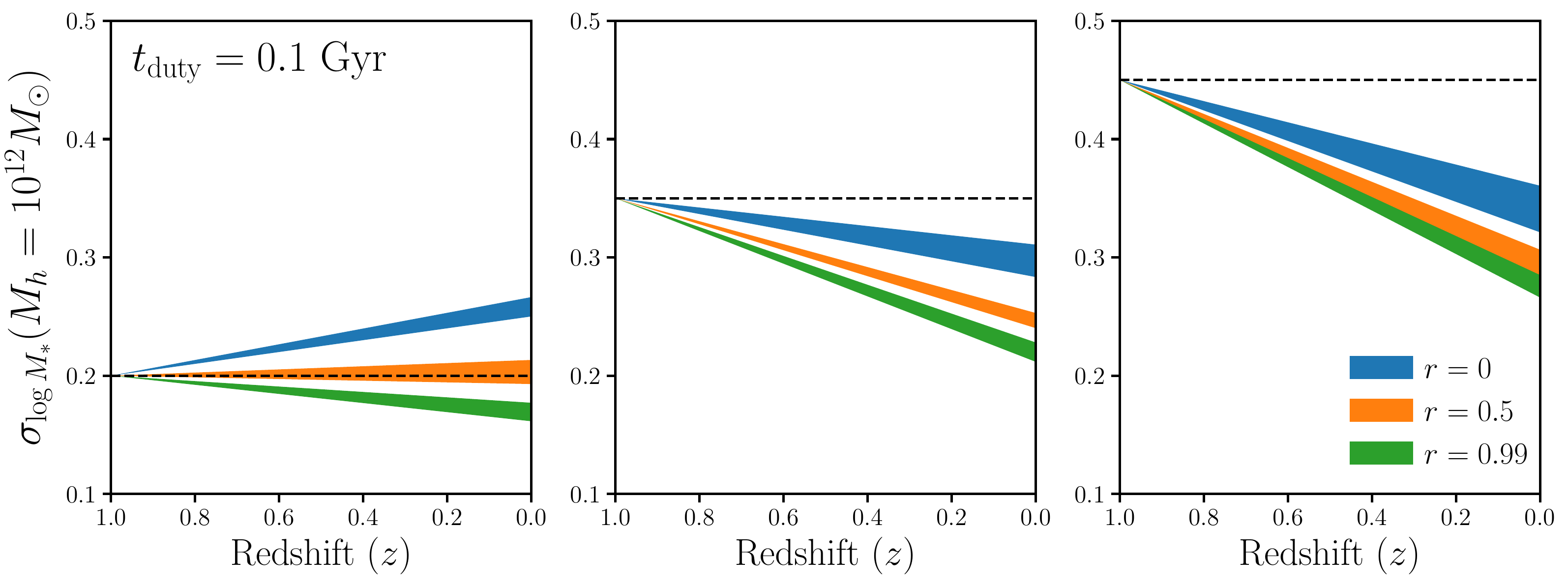}
    \caption{The evolution of $\sigtwe$ from $z=1$ to 0 for our models with 
        $t_{\rm duty} = 0.5$ Gyr and $r = 0$ (blue), 0.5 (orange), and 0.99 (green). 
        In each panel we vary the initial $\sigtwe$ at $z=1$, $\siglogm^{\rm init} = 0.2$, 
        0.35, and 0.45 dex (left, center, and right panels, respectively). 
        $\siglogm^{\rm init} = 0.35$ and 0.45 dex are motivated by $\siglogm$ at 
        $M_h = 10^{12}$ and $10^{11.5}M_\odot$, respectively, in the Illustris TNG 
        (Cao et al in preparation). The width of the shaded region represent the $68\%$ 
        confidence interval. We mark $\siglogm^{\rm init}$ in black dashed. Increasing 
        $\siglogm^{\rm init}$, increases $\sigtwe$ overall. However, for all 
        $\siglogm^{\rm init}$, $r > 0$ reduces the $\sigtwe$ evolution. For the 
        $\siglogm^{\rm init}>0.2$ dex models, even without galaxy assembly bias ($r = 0$), 
        $\tduty = 0.5$ Gyr alone significantly tightens $\sigtwe$ from $z = 1$ to 0 (blue). 
        This is enhanced with $r > 0$. With $r \ge 0.5$, the $\siglogm^{\rm init} > 0.2$ dex 
        models can produce $\sigtwe$ evolutions consistent with the $\sim 0.15$ dex decline 
        in $\sigtwe$ \cao~find in Illustris TNG.}
    \label{fig:siglogM_zevo}
\end{center}
\end{figure}

\subsection{Models with Galaxy Assembly Bias: $r > 0$}
A shorter star formation duty cycle timescale produces tighter scatter in the
SHMR of our fiducial model. This dependence on the duty cycle timescale,
allows us to compare the model to measurements of $\sigtwe$ and predictions 
from galaxy formation models to constrain $t_\mathrm{duty}$, which reflect the 
star formation variability timescale. Such comparisons in the previous section, 
demonstrate that $t_{\rm duty} \lesssim 0.5$ Gyr is favored by observational 
constraints. However, a short duty cycle timescale alone is not enough to 
reproduce $\siglogm$ constraints and its evolutionary trend from halo model 
observations and hydrodynamic simulations. In this section, we examine how 
assembly bias impacts $\siglogm$ using our models that correlate SFH with host 
halo accretion history ($r > 0$).

We repeat our analysis of inferring model parameters by comparing to observations
using ABC-PMC --- this time for our model with galaxy assembly bias over a grid of
$t_{\rm duty}$ and $r$ values. Using the resulting posterior distributions, we
examine $\sigtwe$ predicted by our model as a function of $t_{\rm duty}$ with $r=0$ 
(no assembly bias; blue), $0.5$ (orange), and $0.99$ (green) in 
Figure~\ref{fig:sigMstar_duty_abias}. The shaded regions represent the $68\%$ 
confidence interval of the predicted $\sigtwe$. We again emphasize that for all 
sets of ($t_{\rm duty}$, $r$) our models reproduce the observed SMF of SF centrals 
and SFS. At $t_{\rm duty} \geq 5\,{\rm Gyr}$ we find no significant difference 
in the scatter, regardless of $r$. Below $t_{\rm duty} < 5\,\mathrm{Gyr}$, however, 
$\sigtwe$ of our model decreases significantly as the SFH of SF galaxies are more 
correlated with halo accretion history. For $t_\mathrm{duty} = 0.1$ Gyr, we find
$\sigtwe{=}\,0.26\substack{+0.01\\-0.01},
0.21\substack{+0.01\\-0.01}$, and $0.17\substack{+0.01\\-0.01} $
for $r = 0.0, 0.5$, and $0.99$, respectively.

Comparing our $r > 0$ models to the observational constraints, we find that 
galaxy assembly bias significantly reduces the tensions with halo model observations 
(left panel of Figure~\ref{fig:sigMstar_duty_abias}).
With a short star formation duty cycle ($t_\mathrm{duty} \leq 1\,\mathrm{Gyr}$)
and galaxy assembly bias with $r \ge 0.5$, our model is in agreement with
these $\siglogm \sim 0.2$ dex constraints. On the other hand,
assembly bias increases the tension with
Cao et al. (in preparation), which more specifically constrains $\siglogm$ 
at $M_h\sim 10^{12}M_\odot$. We also compare our $r > 0$ models to
predictions from galaxy formation models on the right panel.
By varying $r$ and $t_\mathrm{duty}$, our model can reproduce the widely varying 
galaxy formation model $\sigtwe$ predictions.
Focusing on hydrodynamic simulations, which best reproduce $\siglogm$
observations at high $M_h$, we find that with a short duty cycle timescale,
$t_\mathrm{duty} < 1$ Gyr, and $r > 0.5$ our models produces the predicted 
$\sim0.2$ dex scatter in SHMR.

A shorter $t_{\rm duty}$ or higher $r$ both produce smaller $\sigtwe$. 
We highlight this degeneracy in Figure~\ref{fig:r_tduty}, where we plot 
$\sigtwe$ (contour and color map) as a function of $t_{\rm duty}$ and $r$. 
Figure~\ref{fig:r_tduty} more precisely reveals that to produce 
$\siglogm \sim 0.2$ dex, $\tduty \leq 1.5$ Gyr and 
$r > 0.7$ (top left corner of Figure~\ref{fig:r_tduty}). 
In the literature, \cite{tinker2018b} find correlation
between $\dot{M_h}$ and $\log$ SSFR with $r = 0.63$ (dashed) and
\cite{behroozi2019} similarly find a correlation between SFH and halo
assembly history with $r_c \sim 0.6$ for halos with $M_h \sim 10^{12}M_\odot$
(dotted). For galaxy assembly bias with $r = 0.6$, the shortest timescale 
we probe ($t_{\rm duty} = 0.1$ Gyr) is necessary to produce 
$\sigtwe \sim 0.2$ dex as found in halo model observations and hydrodynamic 
simulations. This timescale is shorter than the $\sim 0.5$ Gyr timescale that \cite{sparre2015} 
find in Illustris galaxies using a Principal Component Analysis of the SFHs. 
However, it is consistent with the timescales found in the FIRE 
simulations~\citep{hopkins2014, sparre2017}.

We examine the evolutionary trend of $\sigtwe$ in Figure~\ref{fig:siglogM_zevo}, 
where we plot $\sigtwe$ from $z=1$ to 0 for our model with $t_{\rm duty} = 0.5$ Gyr 
and $r = 0$ (blue), 0.5 (orange), and 0.99 (green). The width of the shaded 
region represent the $68\%$ confidence interval. We mark the
initial $\sigtwe$, $\siglogm^{\rm init}$, in black dashed; 
$\siglogm^{\rm init} = 0.2$, 0.35, and 0.45 dex in the left, center, and right 
panels, respectively. Focusing on the $\siglogm^{\rm init} = 0.2$ panel (left),
we find that for $r > 0$ reduces the growth in $\sigtwe$ from $\siglogm^{\rm init}$. 
In fact, for $r = 0.99$, our model (green) predicts $\sigtwe$ that decreases 
from $z=1$. For $r \sim 0.6$~\citep{behroozi2019,tinker2018b}, 
we find a slight growth in $\sigtwe$ from $z=1$: $\sim0.02$ dex (Figures~\ref{fig:r_tduty} 
and~\ref{fig:siglogM_zevo}). This is loosely consistent with halo model observations 
that find constant $\siglogm = 0.2$ dex evolution. However, as we 
discuss in Section~\ref{sec:siglogm_init}, the constant $\siglogm = 0.2$ dex 
evolution from halo model observations is based on $\siglogm^{\rm init} = 0.2$ dex 
constraints in the literature, which do not accurately reflect the SHMR scatter 
at $z \sim 1$ at $M_h = 10^{12}M_\odot$. 

Relaxing the $\siglogm^{\rm init} = 0.2$ dex assumption, we present the 
$\sigtwe$ evolution for our models with $\siglogm^{\rm init} = 0.35$
and 0.45 dex (center and right panels of Figure~\ref{fig:siglogM_zevo}).  
$\siglogm^{\rm init} = 0.35$ and 0.45 dex is motivated by $\siglogm$ at 
$M_h = 10^{12}$ and $10^{11.5}M_\odot$, respectively, in Illustris TNG (Cao et al in preparation). 
Increasing $\siglogm^{\rm init}$, increases $\sigtwe$ overall for all $r$. 
However, both a shorter $t_{\rm duty}$ and higher $r$ produce tighter 
$\sigtwe$ in our $\siglogm^{\rm init} > 0.2$ dex models. Hence, $\sigtwe$ 
remains sensitive to $t_{\rm duty}$ and $r$, regardless of $\siglogm^{\rm init}$. 
Figure~\ref{fig:siglogM_zevo} illustrates that for the $\siglogm^{\rm init}> 0.2$ dex 
models, even without galaxy assembly bias ($r=0$), $t_{\rm duty} = 0.1$ Gyr 
alone significantly tightens $\sigtwe$ from $z=1$ to 0 (blue): $\sim0.06$ and 
$0.11$ dex for $\siglogm^{\rm init} = 0.35$ and $0.45$ dex, respectively. With 
$r > 0$, the decline in $\sigtwe$ from $z=1$ to 0 is further enhanced. With 
$r \ge 0.5$, the $\siglogm^{\rm init}> 0.2$ dex models can produce $\sigtwe$ 
evolutions consistent with the $\sigtwe$ decline \cao~find in the Illustris 
TNG. 

In this section, we use our models with different $\tduty$, $r$, and 
$\siglogm^{\rm init}$ to investigate how these parameters impact predictions 
of $\sigtwe$ at $z{=}0$. A shorter timescale of star formation variability, 
$\tduty$, produces a tighter SHMR scatter. Higher correlation 
between SFH and halo assembly history, higher $r$, also produces a tighter 
SHMR scatter. Furthermore, $\sigtwe$ remains sensitive to $\tduty$ and $r$, 
regardless of $\siglogm^{\rm init}$. Comparing our model predictions to 
$\sigtwe$ constraints in the literature, we find that by varying $\tduty$ 
and $r$ our model can produce $\sigtwe$ loosely consistent with constraints 
from observations and modern galaxy formation models, which span $0.2 - 0.35$ dex. 
To reproduce the constant $\siglogm \sim 0.2$ dex evolution from $z=1$ to 0 
found in halo model based observational constraints, our models require 
$\tduty \leq 1.5$ Gyr for $r = 0.99$ and $r > 0.6$ for $\tduty = 0.1$ Gyr. 
If we fix $r = 0.6$, the constraint on galaxy assembly bias from the literature, 
$\tduty < 0.2$ Gyr is necessary. Meanwhile, to reproduce the 
$\sigtwe \sim 0.35$ to 0.2 dex decline from $z=1$ to 0 found in Illustris TNG, 
our models with $\siglogm^{\rm init} > 0.2$ dex require $r > 0.5$ for $\tduty = 0.1$ Gyr. 
The lack of consensus among observations and galaxy formation models prevents us 
from precisely constraining $\tduty$ or $r$. However, we illustrate that 
$\sigtwe$, the scatter of the SHMR, is sensitive to $\tduty$ and $r$ and, 
thus, demonstrate that measurements of the SHMR relation can be used to constrain 
the detailed star formation histories of SF central galaxies and their connection 
to host halo assembly histories.

\section{Summary and Conclusion} \label{sec:summary}
Despite our progress in understanding how galaxies form and evolve in
the $\Lambda$CDM  hierarchical universe, our understanding of the
detailed star formation histories of galaxies and their connection to host
halo assembly histories have been limited. This is in part due to the
challenges in directly measuring SFHs in both observations and galaxy
formation models. Empirical models, with their flexible prescriptions
have made significant progress in better quantifying the SFHs of galaxies.
These models, however, have yet to examine and constrain the timescale of
star formation variability, which has the potential to constrain physical
processes involved in star formation and galaxy feedback models. In this
paper, we therefore focus on measuring the timescale of star formation 
variability and the connection between star formation and host
halo accretion histories of star-forming central galaxies.

We combine the cosmological $N$-body $\mathtt{TreePM}$ simulation
with SFHs that evolve the SF central galaxies along the SFS and present models
that tracks the SFR, $M_*$, and host halo accretion histories of SF centrals from 
$z \sim 1$ to $z=0.05$. More specifically, we characterize the SFHs to evolve with
respect to the mean $\log\,\mathrm{SFR}$ of the SFS 
and introduce star formation 
variability using a ``star formation duty cycle'', where the SFRs of the SF centrals 
fluctuate about $\logsfrsfs$ on some timescale, $t_\mathrm{duty}$.
We parameterize the SFS using parameters that dictate the low $M_*$ and high $M_*$ 
slopes and redshift evolution. We then compare this model to the observed SMF of the 
SF centrals in the SDSS DR7 group catalog using ABC-PMC likelihood-free 
inference framework. When we examine the SHMR predicted by the model and inferred 
parameters we find:
\bitem
\item A shorter star formation duty cycle in our model produces significantly
    tighter scatter in the SHMR at $M_h = 10^{12} M_\odot$, $\sigtwe$. 
    For $t_\mathrm{duty}$ from $10$ to $0.1$ Gyr, our model predicts
    $\sigtwe{=}\,0.32\substack{+0.019\\ -0.021}$ to $0.26\substack{+0.008\\-0.008}$.
    The dependence of $\siglogm$ on $t_\mathrm{duty}$ demonstrates that
    the scatter in SHMR can be used to constrain $t_\mathrm{duty}$, and thus  
    the timescale of star formation variability.

\item We compare the $\sigtwe$ predicted by our model to observed constraints 
    from halo occupation modeling of galaxy clustering, SMF, satellite kinematics, 
    and galaxy-galaxy weak lensing. There is significant tension among the 
    observational constraints and also among predictions from galaxy formation 
    models with $\sigtwe$ spanning $0.2$ to 0.35 dex. Among the literature, 
    constraints from halo model based observations and hydrodynamic simulations 
    find $\sigtwe \sim 0.2$ dex, which our model struggles to produce even with 
    the shortest timescale we probe, $t_\mathrm{duty} = 0.1$ Gyr.

\item We next examine models with assembly bias that correlate SFHs to host halo 
    accretion histories with correlation coefficient, $r$. With stronger correlation, 
    higher $r$, our models predict tighter scatter in the SHMR down to 
    $\sigtwe{=}0.17$ for $r = 0.99$. To produce $\sigtwe \sim 0.2$ dex, our models 
    require $r > 0.6$ for  $\tduty = 0.1$ Gyr or $\tduty < 2$ Gyr for $r = 0.99$. 
    For $r \sim 0.6$, as found in the literature, $\tduty \lesssim 0.2$ Gyr is 
    necssary. If we allow $\siglogm^{\rm init}$ at $z = 1$ to vary $> 0.2$ dex, we find 
    that our model requires $r > 0.5$ for $\tduty = 0.1$ Gyr to reproduce the
    $\sigtwe = 0.35$ to 0.2 dex evolution from $z=1$ to 0 in Illustris TNG.
\eitem

Our work demonstrates that constraints on the scatter in the SHMR can be
used to constrain both the timescale of star formation variability and
the correlation between SFH and halo accretion history. The main
bottleneck in deriving precise constraints remains the lack of consensus 
among $\sigtwe$ observations both at $z = 0$ and 1.
Also, while we focus on $\siglogm$ at $M_h=10^{12}M_\odot$, the limit 
of current observations, $\siglogm$ is an even more sensitive probe at 
lower $M_h$. Upcoming surveys, however, will make significant progress on 
these fronts. 

The Bright Galaxy Survey of the Dark Energy Spectroscopic Instrument~\citep[DESI;][]{desicollaboration2016}, 
for instance, will observe $\sim$10 million galaxies down to the magnitude 
limit $r \sim 20$ out to $z\sim0.5$. This will enable BGS to to more precisely 
constrain $\siglogm$ and resolve current tensions in observations at 
$z=0$. Meanwhile, the Galaxy Evolution Survey of the Prime Focus Spectrograph~\citep{takada2014,tamura2016}, 
which will observe $\sim500,000$ galaxies between $0.5 < z < 2.0$, and the 
Wide-Area VISTA Extragalactic Survey~\citep[WAVES;][]{driver2016,driver2019}, which 
will observe $\sim$2 million galaxies down to $r_{\rm AB} < 22$ mag out to $z\sim 1$, 
will enable precise constraints of $\siglogm$ at $z\sim1$. 
Using measurements from these surveys, our model will be able to constrain 
the physical processes that govern star formation in galaxies and the detailed
connection between star formation and host halo accretion.

\section*{Acknowledgements}
It's a pleasure to thank
    J.D.~Cohn,
    Shirley~Ho,
    and
    Tjitske~Starkenburg
for valuable discussions and feedback. We also thank Louis E. Abramson,
Junzhi Cao, Shy Genel, and Cheng Li for providing us with data used in
the analysis. This material is based upon work supported by the U.S.
Department of Energy, Office of Science, Office of High Energy Physics,
under contract No. DE-AC02-05CH11231. AW was supported by NASA, through 
ATP grant 80NSSC18K1097 and HST grants GO-14734 and AR-15057 from STScI.

\bibliographystyle{aasjournal}
\bibliography{centralMS}
\end{document}